\documentclass[12pt,preprint]{aastex62}

\usepackage{natbib}
\usepackage{amsmath}
\usepackage{graphicx}
\usepackage{rotating}
\usepackage{enumitem}
\usepackage{url}

\usepackage{pifont}
\newcommand{\cmark}{\ding{51}}%
\newcommand{\xmark}{\ding{55}}%

\begin{document}

\shorttitle{Instabilities in the Early Solar System}
\shortauthors{Quarles \& Kaib}

\title{Instabilities in the Early Solar System due to a Self-gravitating Disk }

\author{B. Quarles \& N. Kaib}
\affil{HL Dodge Department of Physics \& Astronomy, University of Oklahoma, Norman, OK 73019,
USA}
\email{billylquarles@gmail.com}

\begin{abstract}
Modern studies of the early solar system routinely invoke the possibility of an orbital instability among the giant planets triggered by gravitational interactions between the planets and a massive exterior disk of planetesimals.  Previous works have suggested that this instability can be substantially delayed ($\sim$100s Myr) after the formation of the giant planets.  Bodies in the disk are typically treated in a semi-active manner, wherein their gravitational force on the planets is included, but interactions between the planetesimals are ignored.  We perform $N$-body numerical simulations using \texttt{GENGA}, which makes use of GPUs to allow for the inclusion of all gravitational interactions between bodies.  Although our simulated Kuiper belt particles are {more massive} than the probable masses of real primordial Kuiper belt objects, our simulations indicate that the self-stirring of the primordial Kuiper belt is very important to the dynamics of the giant planet instability.  We find that interactions between planetesimals dynamically heat the disk and typically prevent the outer solar system instability from being delayed by more than a few tens of million years after giant planet formation.  Longer delays occur in a small fraction of systems that have at least 3.5 AU gaps between the planets and planetesimal disk. Our final planetary configurations match the solar system at a rate consistent with other previous works in most regards.  Pre-instability heating of the disk typically yields final Jovian eccentricities comparable to the modern solar system value, which has been a difficult constraint to match in past works. 
\end{abstract}

\section{Introduction}
Over the last few decades, it has become clear that the solar system's outer planets have evolved substantially from their initial orbital configuration \citep{Fernandez1984,Malhotra1993}.  While evidence of this process mostly comes from the observed features of bodies orbiting beyond Neptune, the effects of giant planet migration would have been felt throughout the solar system.  One particularly successful realization of giant planet migration known as the ``Nice Model'' assumes that the gas giants were able to form in a different  configuration before dissipation of the primeval gaseous disk \citep{Gomes2005,Tsiganis2005}.  However, the resulting system of giant planets was only meta-stable, where an instability caused these planets to scatter into the orbital configuration we see today \citep{Gomes2005,Tsiganis2005,Morbidelli2007a,Morbidelli2009}.

The Nice model has been used to explain different dynamical phenomena in the solar system, such as Jupiter's Trojans \citep{Morbidelli2005,Robutel2006}, the capture of irregular satellites \citep{Cuk2006,Nesvorny2007,Jewitt2007}, the dynamical structure of the asteroid belt \citep{Morbidelli2009,Minton2011,Roig2015}, and a possible dynamical trigger for a Late Heavy Bombardment \citep{Bottke2007,Morbidelli2007b,Thommes2008,Bottke2012}.  The initiation of the instability and the scattering of small bodies that followed has been linked to an epoch of bombardment on the Moon that has been recorded within its craters (see \cite{Bottke2017} for a review).  This late period of bombardment is thought to set the timescale for the giant planets to scatter at $\sim$500 -- 700 million years (Myr) after the condensation of Calcium-Aluminum rich inclusions in the early protoplanetary disk \citep{Gomes2005,Tsiganis2005}, although recent studies \citep{Toliou2016,Deienno2017,Zellner2017,Morbidelli2018,Clement2018} have begun to call into question this assumption.

Due to the broad range of masses  (i.e., asteroids, comets, embryos, and planets), numerical studies have often used approximations in order to make the computations more tractable.  For instance, the outer debris disk is usually described as swarms of real disk particles that either interact only with the gas giants ignoring the gravitational interactions between swarms, or allow interactions between the swarms, but only within a few mutual Hill spheres to approximate the effects of viscous self-stirring \citep{Gomes2005,Tsiganis2005,Levison2011}.  {The latter approach was used by \cite{Levison2011}, where they assumed that most of the self-heating in the outer disk should have come from encounters between small particles in the disk that represent the bulk of its mass and $\sim$1000 Pluto-sized particles, which represented the 'heavy' part of the size distribution.  That method showed to be computationally efficient and self-stirring was shown to be small but not negligible.  For a large enough spacing between the outer gas giant and the inner edge of the disk, there were practically no encounters between particles and planets for very long times due to negligible disk spreading in semimajor axis.  Self-stirring was found to secularly transfer energy to the giant planets, which eventually became unstable and constituted a regime of solutions, all necessarily late.}

Within modeling efforts of these types, there has also been work on the number of giant planets present within this era of solar system evolution. Such works have explored how the existence of 4, 5, or 6 giant planets would affect the final orbital architectures, the migration rate, and the instability timescale necessary to match observational constraints in the Asteroid Belt and the craters of the inner solar system \citep{Tera1974,Michel2007,Bottke2012,Morbidelli2017}. In particular, recent studies have shown that a giant planet instability very often destabilizes the terrestrial planets, again calling into question the timing of instability \citep{Brasser2009,Nesvorny2011,Nesvorny2012,Agnor2012,Brasser2013,Kaib2016}. A less delayed instability occurring before the completion of terrestrial planet formation may be more compatible with the present-day solar system \citep{Clement2018,Nesvorny2018}.  

{\cite{Fan2017} used a self-gravitating disk model and found consistent final giant planet architectures, but did not evaluate the full timescales associated with the Nice Model.}  In this work, we employ models with a self-interacting disk to revisit the timing and outcomes of an outer solar system instability on 100 Myr timescales.  {Through these models we estimate the magnitude of excitation within the disk from three sources: external forcing by the giant planets, self-stirring due to particle-particle interactions, and artificial enhancement of the self-stirring due to mass resolution of particles.}  Our initial setup and methodology is summarized in Section \ref{sec:methods}.  We describe our results for special cases in Sections \ref{sec:short_term} and \ref{sec:isolated}, where our broader results for 4 and 5 giant planet systems are presented in Sections \ref{sec:4GP} and \ref{sec:5GP}, respectively.  We discuss the dependance of our results on the number of particles assumed in Section \ref{sec:num_part} and \ref{sec:hires}. The conclusions of our study are summarized in Section \ref{sec:conc}.

\section{Methodology} \label{sec:methods}
\subsection{Numerical Setup}
Our numerical study makes use of a relatively new code based upon the popular integration package \texttt{mercury} \citep{Chambers1999} that has been adapted for GPUs called \texttt{GENGA} \citep{Grimm2014}.  This code has been well-tested and shown to yield results consistent with the \texttt{mercury} package \citep{Grimm2014,Hoffmann2017}.  In our study, we use NVIDIA Tesla K20m cards with a compute capability of 3.5 and version 8.0 of the CUDA drivers.  In practice, the code allows for 3 modes of operation: fully-active, semi-active, and test particles, similar to the standard version of \texttt{mercury}.  The test particle mode allows large bodies (fully-active particles) to influence the motions of massless bodies (test particles), ignoring the interactions between test particles and the reaction forces onto the massive bodies.  The semi-active mode upgrades the test particles to small bodies that have mass and allows the reaction forces on the large bodies, but still ignores the interactions between small bodies.  We primarily use the fully-active mode, where all gravitational interactions are included between all bodies and use the other modes for studies of special cases.  

Most of our simulations begin with a specified giant planet resonant configuration and an outer disk of 1500 smaller bodies.  The outer disk is composed of equal mass bodies distributed following a surface density profile $\Sigma \propto a^{-1}$, which is common among previous investigations \citep[See][]{Levison2011}.  We vary the inner edge of the disk $a_i$, or disk gap ($\Delta = a_i - a_o^{GP}$), following previous works within $\sim$2.3 -- 6.3 AU from the outer most giant planet, $a_o^{GP}$, in increments of 0.125 AU at our highest resolution.  The outer boundary of the disk is kept initially fixed at 30 AU.  However we expect spreading due to interactions within the disk that push bodies beyond this boundary.  We use 100 AU as a radial boundary to consider bodies to be ejected and  bodies that extend beyond this boundary are removed from a given simulation.  We perform a subset of runs using this basic setup, where we primarily vary the number and/or mass of the small bodies to measure the extent of the self-stirring using the root-mean-squared eccentricity of disk particles on a 1 Myr timescale.

\begin{figure}
\begin{center}
\includegraphics[width=\textwidth]{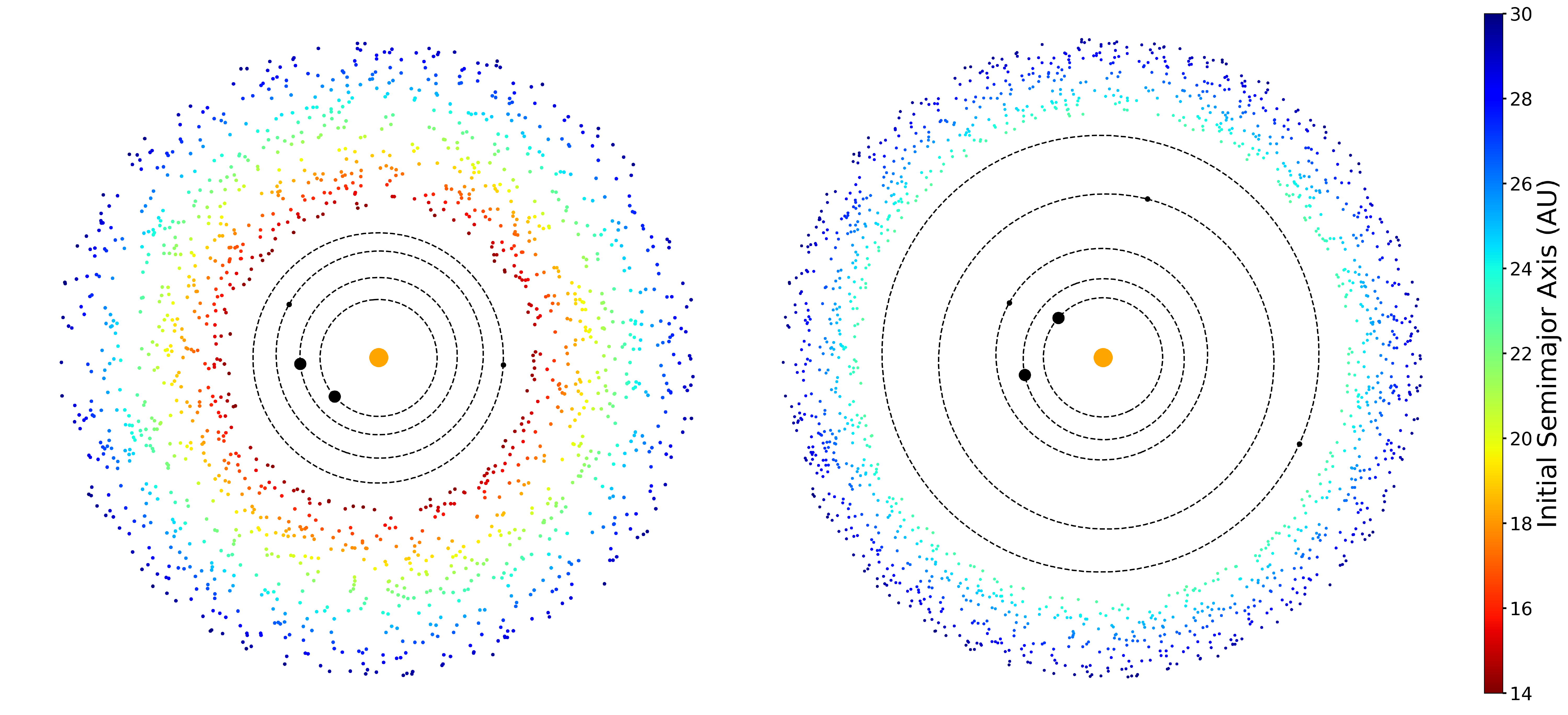}
\end{center}
\caption{Representations of the initial system state in the x-y plane considering a Nice Model configuration with 4 giant planets (left) with a 35 M$_\oplus$ outer disk ($a_i = 14$ AU) or with 5 giant planets (right) with a 20 M$_\oplus$ outer disk ($a_i = 22.697$ AU).}
\label{fig:orbits_IC}
\end{figure}

A large fraction of our simulations are terminated when 50 Myr of simulation time has elapsed, while a subset of runs were extended beyond 50 Myr.  These runs where chosen because they contained a significant portion of their disk at 50 Myr and had at least 4 giant planets for a delayed instability scenario to remain plausible.  We use a timestep of 180 days for our runs, which is typical within previous works \citep{Levison2011,Nesvorny2012}.  The initial eccentricities and inclinations of the disk particles are chosen randomly from a Rayleigh distribution with a scale parameter $\sigma = 0.001$.  The other orbital parameters (argument of periastron, ascending node, and mean anomaly) are chosen randomly from a continuous uniform distribution between 0--360$^\circ$.  Figure \ref{fig:orbits_IC} shows the initial state of our simulations from a top-down perspective for both the four giant planet (left) and five giant planet (right) configuration.

\subsection{Criteria for Success}\label{sec:criteria}
Following \cite{Nesvorny2012}, we prescribe 4 criteria to measure the overall success of our simulations.  These criteria are outlined as follows:

\begin{enumerate}[label=(\Alph*)]
\item The final planetary system must have 4 giant planets,
\item The final orbits of the giant planets must resemble the current solar system,
\item The $e_{55}$ amplitude must be greater than half of its current value (0.044), and
\item The period ratio between Jupiter and Saturn changes from $<2.1$ to $>2.3$ within 1 Myr.
\end{enumerate}

Our criteria are virtually identical to those of \cite{Nesvorny2012} so that we can fairly compare our results to other previous investigations \citep[i.e.,][]{Levison2011,Nesvorny2012}.  Similar to \cite{Nesvorny2012}, we expect that only a few percent of systems will satisfy all 4 criteria.  The justification of Criterion A is self-evident when comparing to the actual solar system.  Criterion B is defined further to include semimajor axes within 20\% of the current values of the giant planets, mean final eccentricities of each planet $<$ 0.11, and mean final inclinations of each planet $<$ 2$^\circ$.  {The mean final eccentricities and inclinations of each planet} are determined from an additional integration after the instability occurs in isolation (without disk particles) for 10 Myr.  Numerical simulations of this type are susceptible to the inherent chaos within dynamical systems resulting in mainly statistical comparisons and from this we justify a fairly wide range in Criterion B.  

Criterion C was previously justified because the eccentricity modes of the giant planets were relatively hard to excite given that the self-gravity of the disk was largely ignored.  We keep this criterion in our analysis, but find that the $e_{55}$ amplitude to be excited relatively easily.  In order to determine the value of $e_{55}$, we perform an integration for 10 Myr using the final giant planet configurations and discard any remaining disk bodies.  This secondary integration is analyzed using the Frequency Modified Fourier Transform\footnote{\url{https://www.boulder.swri.edu/~davidn/fmft/fmft.html}.} by \cite{Sidlichovsky1996}.  The secular mode of Jupiter has been used to broadly describe the long-term evolution of the solar system as it affects the observed structure found in populations of small bodies.  Criterion D institutes a quick transition through resonances that would sweep through the terrestrial region of the solar system \citep{Brasser2009,Kaib2016}.  We keep Criterion D in our analysis for completeness, but do not strictly adhere to it when describing simulations as successful or not because its importance is diminished if the instability occurs before the epoch of terrestrial planet formation is complete \citep{Clement2018}.

\subsection{Giant Planet Architectures}
{Previous studies have used giant planet architectures, where the period ratios ($T_{out}:T_{in}$) between successive planets are integer ratios $a:b$ that correspond to the mean motion resonances between the giant planets expanding radially outward \citep[e.g.,][]{Morbidelli2007b,Batygin2010}.  A wide range of initial configurations have been proposed where several authors start Jupiter and Saturn near the 2:1 MMR, as it is the strongest resonance \citep{Tsiganis2005,Morbidelli2007b,Zhang2010, Pierens2014, Izidoro2016}.  However, \citet{Nesvorny2012} were unable to sufficiently excite $e_{55}$ with this starting condition, so we restrict our initial conditions to resonant chains where Jupiter and Saturn occupy a 3:2 MMR.}  
 
Our simulations use architectures of 4 and 5 giant planets that have been shown to be successful in prior investigations \citep{Levison2011,Nesvorny2012}.  The system of 4 giant planets considers Jupiter and Saturn as fully grown planets and 2 ice giants (15.75 M$_\oplus$ each) in a 3:2, 3:2, 4:3 resonant configuration.  The system of 5 giant planets follows from a 3:2, 3:2, 2:1, 3:2 resonant configuration that was shown to be particularly good for a delayed instability \citep{Nesvorny2012,Deienno2017}.  Using \texttt{GENGA}, we find that both of these resonant configurations are stable up to 600 Myr without including any disk particles.  

\section{Results and Discussion}

\subsection{Short Term Evolution of an Isolated Disk} \label{sec:short_term}
In order to make fair comparisons with previous works, we performed a series of simulations for 1 Myr using similar initial conditions as \cite{Levison2011} for 4 giant planets and a 5 giant planet configuration from \cite{Nesvorny2012}.  The particles used {in} most of our runs are more massive than the largest known KBOs, and thus it is important to understand how our assumed mass resolution of particles and disk surface density will affect the outcomes in our simulations.  In particular, we want to estimate how much disk-stirring is enhanced by using super-Pluto mass particles. To do this, we can compare the behavior of short fully interacting simulations with similar ones done with Pluto-mass bodies in past work (Levison et al. 2011).  From these simulations, we measure the viscous stirring of the disk particles through the root-mean-squared eccentricity, $e_{\rm rms}$.   

\begin{figure}
\begin{center}
\includegraphics[width=\textwidth]{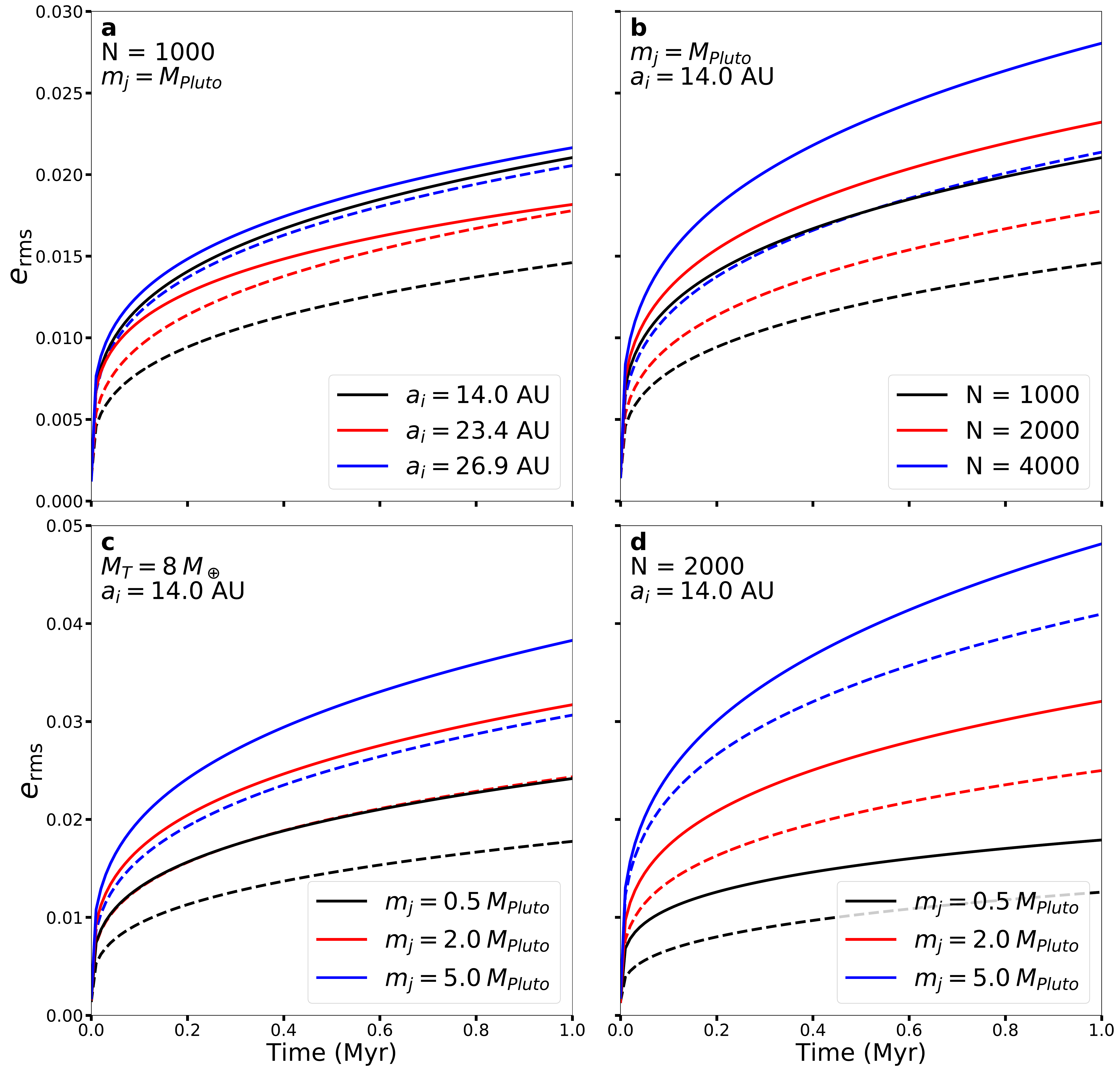}
\end{center}
\caption{Measures of the viscous stirring in terms of the $e_{\rm rms}$  considering a Nice Model configuration with 4 giant planets, where the outer disk has: a) 1000 Pluto-mass bodies with 3 values for the inner edge of the disk ($a_i$), b) a constant surface area, but a different  number of Pluto-mass bodies, c) a constant total mass and area using 3 particle masses, and d) a constant surface area and number of bodies using 3 particle masses.  The colors distinguish between different values ($a_i$, particle number, and particle mass), where we include simulations with (solid) and without (dashed) the giant planets.}
\label{fig:stir4}
\end{figure}

\begin{figure}
\begin{center}
\includegraphics[width=\textwidth]{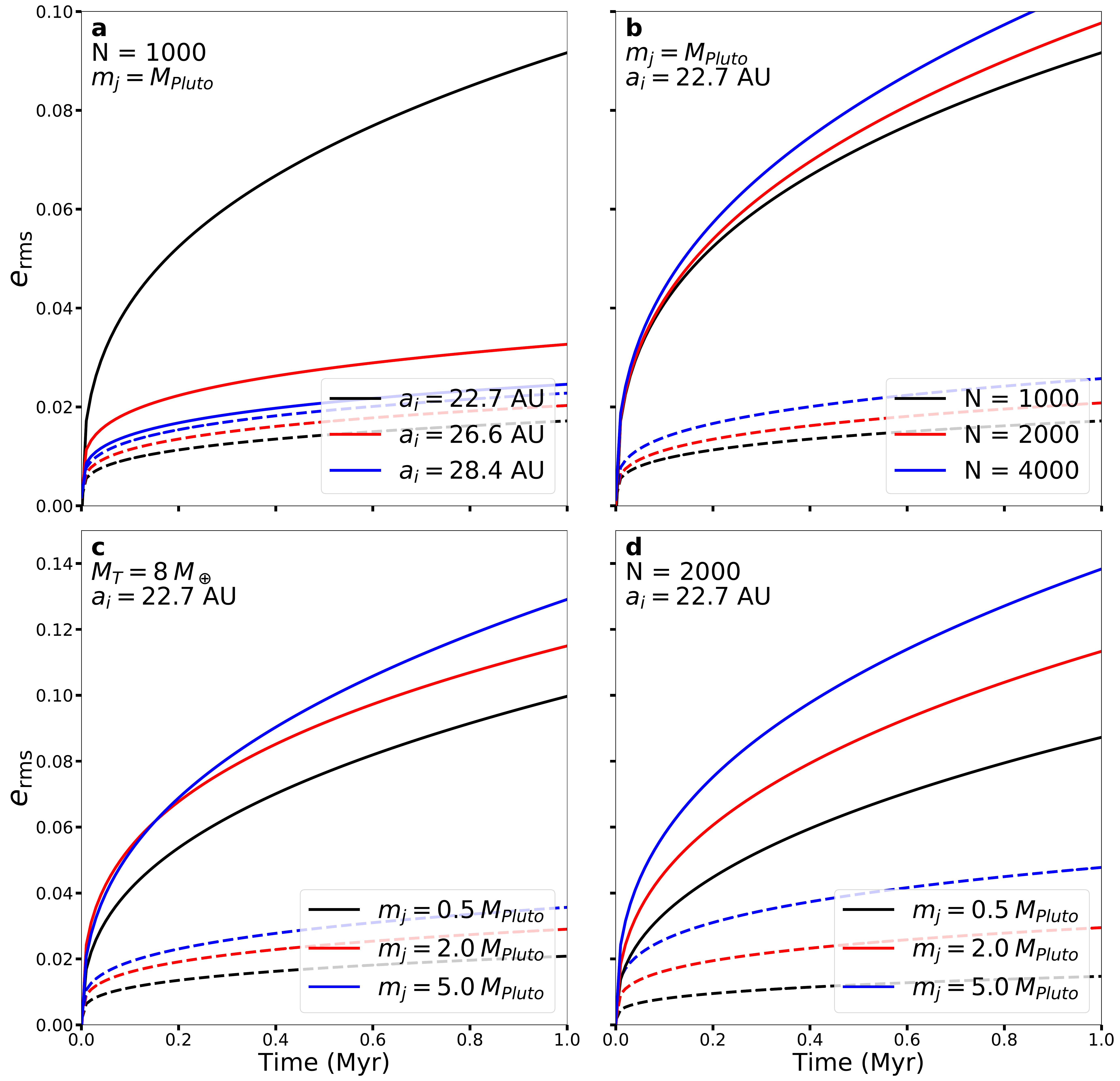}
\end{center}
\caption{Similar to Figure \ref{fig:stir4}, but using a Nice Model configuration with 5 giant planets.}
\label{fig:stir5}
\end{figure}

First, we evaluate the short term evolution of an outer disk using our 4 and 5 giant planet configurations that begin in a resonant configuration in Figures \ref{fig:stir4} and \ref{fig:stir5}, respectively, where we systematically explore 4 different scenarios.  In each scenario, we evaluate the prescribed conditions with and without the giant planets present so that we can more easily separate the  interactions of the disk bodies from either the self-stirring of the disk or the secular forcing from the inner giant planets.

The first scenario in Fig. \ref{fig:stir4}a starts with an outer disk composed of 1000 Pluto-mass bodies, where the initial surface density ($\sigma_D$) of the disk varies between simulations.  We keep the total mass constant and change the inner edge of the disk ($a_i$) while keeping the outer edge of the disk fixed at 30 AU.  Beginning the inner edge of the disk at $\sim$23.4 or 26.9 AU results in doubling or quadrupling, respectively, the initial surface density, $\sigma_D$, compared to setting the inner edge at 14 AU.  As one would expect, the viscous self-stirring increases {in proportion to} increases in $\sigma_D$ for isolated disks\citep[e.g.,][]{Stewart2000,Levison2011}, where our simulations (dashed lines) indicate self-stirring between small bodies contribute a significant fraction to the total viscous stirring of the disk.  In particular, we find good agreement comparing to previous works \citep[][see their Figure 2]{Levison2011} that find $e_{\rm rms}\sim 0.02$ after 1 Myr when the inner edge of the disk is 14 AU.

{In Figure \ref{fig:stir5}a we perform the same experiment on our 5 planet configuration. Here we see that the self-stirring in isolated disks (dashed lines) is comparable to Fig. \ref{fig:stir4}a. However, when we introduce the giant planets (solid lines) into the system, the disk is driven to much higher eccentricities compared to the corresponding 4 planet simulations with a similar disk gap. This appears to be driven by planet-disk interactions.  Previous work \citep{Levison2011} has shown that planet-disk interactions are strongest for portions of the disk within or closer to the 3:2 resonance with the outer giant.  In our 4 planet setup, this region of the disk hosts only $\sim$4\% of our particles (assuming a $\sim$3 AU gap between the disk and planets). Meanwhile, 51\% of disk particles are found in this region if we assume the same planet-disk gap in our 5 planet configuration! Thus, a much larger fraction of the disk is strongly influenced by the giant planets.  One can also see this in the red and blue solid curves in Fig. \ref{fig:stir5}a where we, respectively, move the inner edge of the disk to 26.6 and 28.4 AU, just beyond the location of the 3:2 MMR. In these disks, the growth in eccentricity is greatly diminished compared to the disk with an inner edge at 22.7 AU. }

In the second scenario (Figs. \ref{fig:stir4}b \& \ref{fig:stir5}b), we keep the surface area of the disk fixed and vary the number of Pluto-mass bodies (1000-4000) present to simulate more realistic conditions \citep[i.e.,][]{Nesvorny2016}.  We find that the viscous self-stirring increases with the number of Pluto-mass bodies included, which is likely due to the relative strength and frequency of close encounters between particles.  \cite{Levison2011} performed simulations that employed 1000 Pluto-mass bodies, but more recent studies of Kuiper belt formation indicate that the primordial belt may have contained as many as 4000 Pluto-mass bodies \citep{Nesvorny2016}.  When we introduce the planets to our simulations in our 4 planet systems (Fig. \ref{fig:stir4}b), the RMS eccentricity {increases by $\sim$0.005-0.01, where the increase is smaller ($\sim$0.005) for more particles ($N=4000$) and vice versa.}  However, our 5 planet simulations (Fig. \ref{fig:stir5}b) show the effects of particle number variations are dwarfed by the eccentricity increases driven by the planets.  In these cases, the RMS eccentricity increases by $\sim$0.1.

In our third set of experiments (Figs. \ref{fig:stir4}c \& \ref{fig:stir5}c), we keep the total surface density of our disks fixed and vary the mass resolution of individual particles (0.5 -- 5.0 M$_{Pluto}$).  This allows us to isolate how the level of disk self-stirring will be impacted by how well we can resolve the mass of the primordial disk.  As a result of varying particle mass over an order of magnitude, we find the viscous stirring variations (as measured by $e_{\rm rms}$) to be similar to that seen when varying the number of Pluto-mass bodies between 1000 and 4000, which spans the actual range of uncertainty of this number.  In fact, the difference in viscous stirring in our isolated disks of 5.0 M$_{Pluto}$ is less than a factor of two larger than if 0.5 M$_{Pluto}$ bodies were used.  The larger effect arises when we use 4 (Fig. \ref{fig:stir4}c) or much more so with 5 (Fig. \ref{fig:stir5}c) giant planets. { The effect in the 5 giant planet scenario is important because it has been shown to better reproduce the current solar system architecture \citep[i.e.,][]{Nesvorny2012}.}

The final scenario (Figs. \ref{fig:stir4}d \& \ref{fig:stir5}d) investigates how the viscous stirring changes with the mass resolution of particles, but the number of particles (2000) and surface area of the outer disk remain fixed.  Increasing particle mass here lowers the mass resolution and increases the disk surface density, both of which lead to increased disk stirring.  As expected, increasing the particle mass drives increases in the disk $e_{\rm rms}$, but as we have seen in our other sets of experiments with our 5 planet configuration, the particle/disk-dependent eccentricities are smaller than the increase in eccentricity that occurs when we introduce the giant planets to the system.

{In each of our short-term experiments we see that the eccentricity stirring is affected by particle number, particle mass, disk surface density, and the perturbations of the giant planets.  In particular, we find that the largest increases in the $e_{\rm rms}$ occur when we embed our 5 planet configuration interior to the primordial disk.  This effect dominates over all of our variations in disk properties. Much of this is due to the fact that a much larger fraction of the disk resides closer to the Sun than the 3:2 MMR with the outer giant.  In addition, the eccentricity evolution of the 4 and 5 giant planets in isolation for 1 Myr shows that the eccentricity of the outer ice giants are larger (up to 2$\times$) in the 5 giant planet system.  These higher planetary eccentricities widen resonances and increase secular forcing, which add to the larger eccentricity growth seen in disks surrounding our 5-planet configuration.  This is important because our longer-term simulations use particle masses that are typically a factor of a few greater than Pluto, and these shorter simulations indicate that our computationally driven choice of larger particle masses should not dramatically alter the evolution of our disks since their stirring is primarily enhanced by the inclusion of the 5 planets.  It is also possible that some of the enhanced eccentricity growth of the 5 planet system is due to the inner edge of the disk being very close to the planets, where particles at the inner edge of the disk get scattered by the planets to high eccentricities very fast (or gain eccentricity in the resonances with the outer giant).  Hence the $e_{\rm rms}$ increases faster than it would in a disk with a significant source of dynamical friction (i.e., dust) creating some artificial enhancement to the growth of the disk $e_{\rm rms}$.}

\subsection{Longer Term Evolution of an Isolated Disk} \label{sec:isolated}
{We perform a number of simulations to evaluate the extent of self-stirring within isolated disks (35 \& 20 M$_\oplus$) using equal mass planetesimals on longer timescales (10 -- 100 Myr).  Longer simulation timescales also translate into longer computational times, where we limit the scope of our numerical tests in response. }

\begin{figure}
\begin{center}
\includegraphics[width=\textwidth]{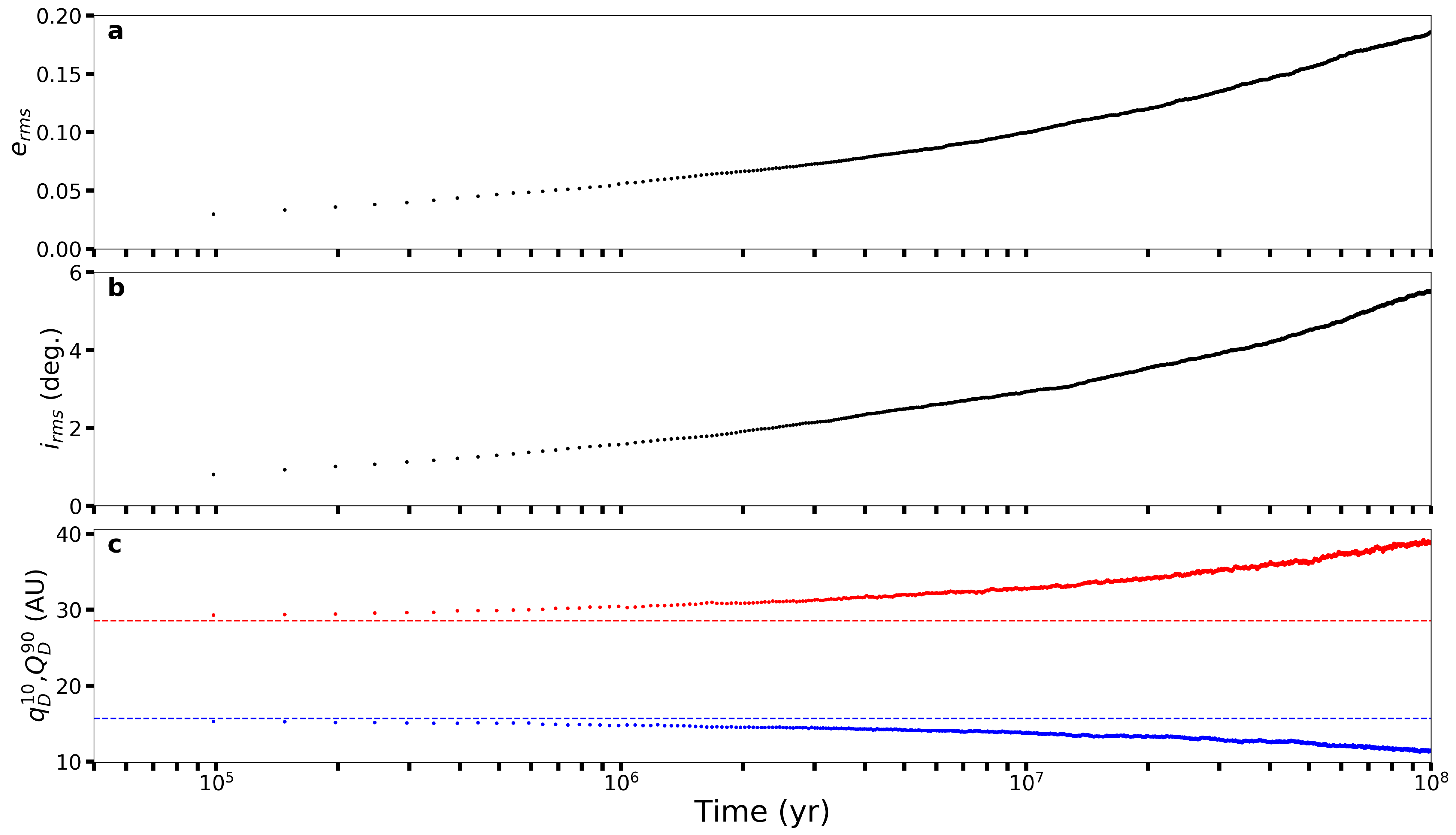}
\end{center}
\caption{Evolution of the (a) rms eccentricity, (b) rms inclination, and (c) the approximate borders of an isolated disk ($a_i = 14$ AU, $M_D = 35 M_\oplus$) without any giant planets.  Panel c delineates the borders of the disk as the 10$^{th}$ percentile periastron $q^{10}_D$ (blue) and the 90$^{th}$ percentile apastron $Q^{90}_D$ (red), where the respective initial values are marked with horizontal dashed lines.}
\label{fig:disk}
\end{figure}

{First, we show the evolution of a 35 M$_\oplus$ disk ($a_i = 14$ AU) composed of 1500 bodies ($\sim$11 Pluto Masses) in terms of the rms eccentricity and rms inclination in Figures \ref{fig:disk}a and \ref{fig:disk}b, respectively.  The $e_{rms}$ and $i_{rms}$ of the disk particles increase exponentially in simulation time, where doubling occurs roughly with an order of magnitude longer simulations.  The magnitude of the increase is likely related to our mass resolution, the infrequent nature of collision events, and the propensity for bodies to scatter off each other.  The 10$^{\rm th}$ percentile of the disk particles in their periastron distances ($q^{10}_D$, blue) and the 90$^{\rm th}$ percentile in their apastron distances ($Q^{90}_D$, red) are also provided in Fig. \ref{fig:disk}c to demonstrate the extent of the disk spreading through the difference in their initial values (dashed, horizontal lines).  The spreading in these distances indicates that the self-stirring of the disk particles will interact with the giant planets eventually and induce the migration of the outermost giant planet through scattering events.}

\begin{figure}
\begin{center}
\includegraphics[width=\textwidth]{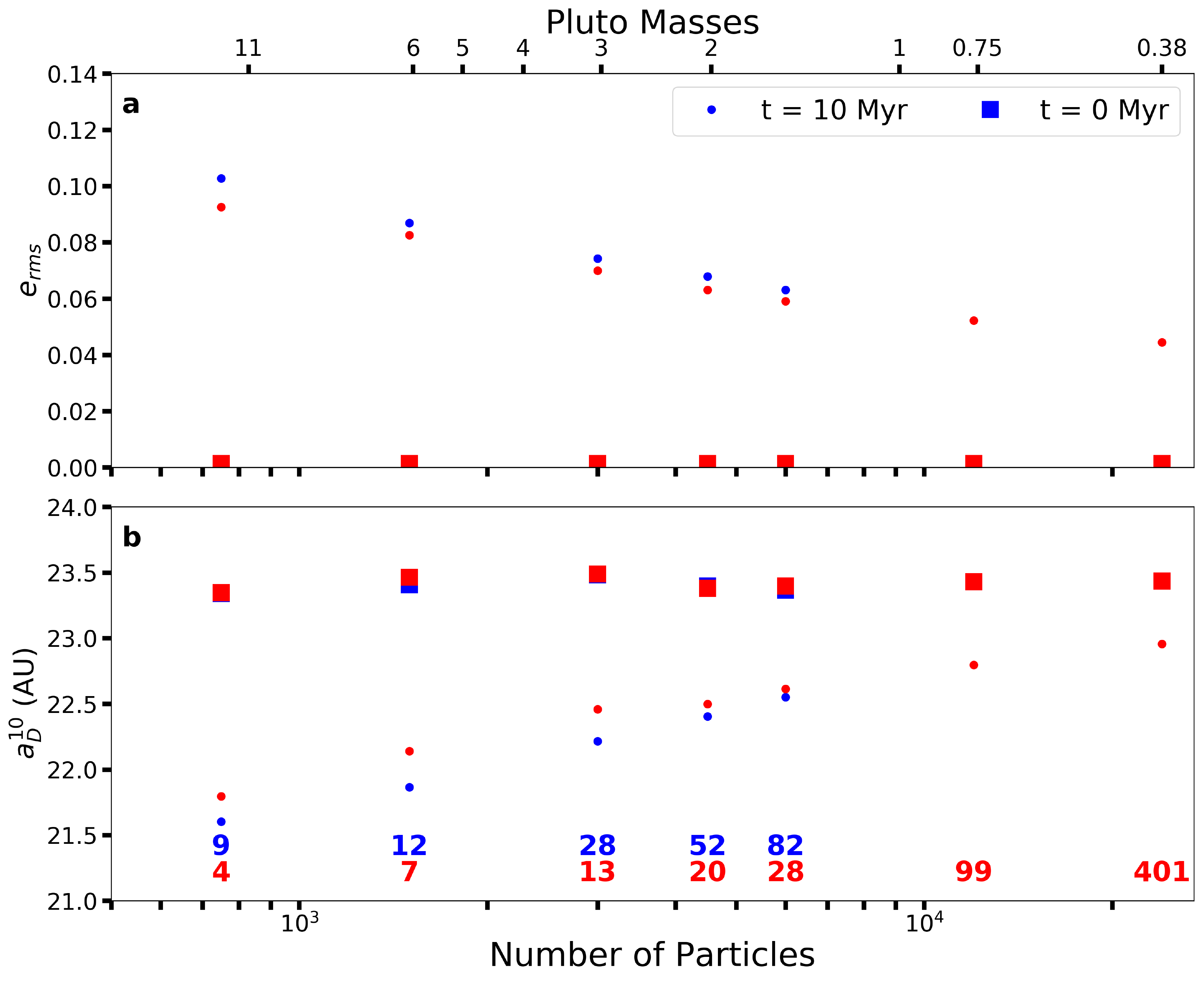}
\end{center}
\caption{{Measures of the viscous stirring with respect to the mass resolution and as a function of time using the (a) $e_{\rm rms}$ and (b) the approximate inner border of an isolated disk using the 10$^{th}$ percentile in the semimajor axis distribution at the initial time (squares) and after 10 Myr of simulation time (dots). The blue symbols use a full disk with denoted number of particles, where red symbols denote simulations where only the inner half of the disk (by area) is evolved.  The color coded values noted in panel b refer to the wall time for each simulation in hours. Note: the difference between measures of full disks (blue) and disks evolving only the inner half (red) rapidly become small when the number of particles increases.}}
\label{fig:isodisk}
\end{figure}

{Mass resolution undoubtedly plays a role in the excitation of the disk particles and can contribute to an enhancement to the actual self-stirring within a disk, where more realistic conditions \citep{Nesvorny2016} would offset this enhancement, but then a large number of particles would be required.  To investigate the potential effects on our results, we simulate isolated 20 M$_\oplus$ disks varying the number (750--24000) of equal-mass planetesimals with an inner edge of the disk beginning at 22.7 AU for 10 Myr.  In Figure \ref{fig:isodisk}, we show the excitation of the eccentricity and the drift of the inner portion of the disk through the 10$^{th}$ percentile of semimajor axis distances.  The symbols in Fig. \ref{fig:isodisk} denote the respective values at two different epochs, where the color code denotes simulations using either the full extent of the disk (blue) or only the inner half (by area) of a disk with an equivalent mass resolution (red).  Probing mass resolution in this way is important because it can play a role in influencing the outer giant through semimajor axis diffusion of the inner disk and the outer half of the disk contributes on much longer timescales due to the increase in orbital period.  }

{Even for 10 Myr integration times, simulating disks with several thousand particles results in very long run times.  Thus, for the highest particle tests (12,000 \& 24,000), we instead only simulate the inner half of the disk, which just consists of 6,000 or 12,000 particles, respectively.  Our justification for this approach is that any given particle in an isolated disk is primarily stirred via interactions with other nearby particles, so the influence of the outer half of the disk on stirring of the inner half is likely small on short ($\lesssim$10 Myr) timescales.  To verify this, we simulate full  disks (up to 6,000 particles) for 10 Myr and then repeat the simulation using only the inner half of the disk resolved with half as many particles.  Examining Figure \ref{fig:isodisk}a, we see that the rms values of disk eccentricity are nearly identical for both systems consisting of more than 1500 particles after 10 Myr. In addition, we see that the inner disk edge (as measured by the 10$^{th}$ pericentile of the semimajor axis distribution) has diffused nearly the same amount when the number of particles is larger than 3000.  With these results, we can effectively study the evolution of a 12,000 or 24,000 particle disks by modeling just the inner half with 6,000 or 12,000 particles, respectively. }

{Figure \ref{fig:isodisk}a shows the $e_{\rm rms}$ after 10 Myr of simulation time for disks composed of relatively large equal-mass bodies ($\sim$11 Pluto Masses) down to a much smaller mass resolution ($\sim$0.4 Pluto Masses), where there is factor of $\sim$2-3 (see red dots in Fig. \ref{fig:isodisk}a) in self-stirring between these mass scales.  KBO surveys indicate that the large KBO size distribution follows a $q\sim{-5}$ size distribution \citep{Shankman2013,Fraser2014,Lawler2018}.  {If we assume there were originally 4000 primordial objects more massive than Pluto and extend such a size distribution to lower masses, we would estimate that there were $\sim$6000 primordial belt objects with masses above 0.75 Pluto Masses.}  This is within a factor of 2 of our 12,000 particle simulation.  As we saw in the Section \ref{sec:short_term}, changing particle number by a factor of two results in modest changes, especially compared to the effects of planet-disk interactions.  Moreover, the average particle mass in our highest resolution simulation ($\sim$ 0.4 Pluto Masses) is less than the average mass of all bodies larger than 0.75 Pluto Masses under the $q\sim{-5}$ distribution, and our underestimate of mass will somewhat offset our overestimate of objects.  Thus, the behavior of this simulated disk, which is not radically different from our coarser isolated disks, should resemble the behavior of the high-mass objects in the actual primordial belt.}  {We note that smoothly extending such a distribution down to $\sim$100-km bodies implies an unrealistically massive primordial belt.  We just employ this example to illustrate that our simulation resolution approaches physical values, given the uncertainty in the primordial belt's properties.}

{While eccentricity stirring can cause disk particles to strongly interact with planets and trigger an instability, the giant planets can also be destabilized through energy diffusion in the disk, which causes the inner disk edge to bleed inward toward the planets.  Thus, we also study the evolution of the 10$^{th}$ percentile of the semimajor axis distribution (as a proxy for its inner edge) in Fig. \ref{fig:isodisk}b.  The diffusion in the 10$^{th}$ percentile semimajor axis distribution for disks composed of a smaller number of more massive particles ($\Delta a_{10} \sim 1.5$ AU) is greater than disks composed of a larger number of less massive particles ($\Delta a_{10} \sim 0.5$ AU).  Our simulations with giant planets mitigate this enhanced diffusion, due to mass resolution, by starting with a gap ($\sim$2.4 AU) between the disk and the outer giant.  With such a gap, diffusion due to self-stirring should take $\sim$100 Myr to begin crossing orbits with the outermost giant planet considering particle masses of $\lesssim$6 times that of Pluto.  Secular forcing due to the giant planets will likely shorten this timescale, where larger gap sizes will counteract the effect.}

{There is a substantial computational cost to consider disks with a larger number of less massive particles, where compute times become intractable using  Pluto to sub-Pluto mass resolutions for our current study.  As a result, the focus of our work is limited to 4500 particles, or $\sim$2 Pluto Masses.  With a more realistic mass resolution (24,000 particles), the $e_{rms}$ value is $45\%$ smaller than with 4500 particles, but secular forcing of eccentricity due to the 5 giant planets will likely overshadow this difference (see Fig. \ref{fig:stir5}c).  In the case of 4 giant planets (see Fig. \ref{fig:stir4}c) the secular forcing is not as large as for 5 giant planets, but such forcing will likely dominate over the enhanced self-stirring due to mass resolution because the surface density $\sigma_D$  in our simulations with 4 giant planets is about half as much as the runs with 5 giant planets.}

\subsection{4 Giant Planets} \label{sec:4GP}

The classical Nice model posits that the four giant planets evolved into a mean motion resonance before the dispersion of the gaseous disk and later transitioned to the current configuration \citep{Gomes2005,Morbidelli2007a}.  {Although most recent research has focused on a 5-planet initial configuration, we first explore the effects of a self-gravitating disk on 4-planet setups for completeness and comparison with past work.}  Our simulations explore how one of the best 4 giant planet cases (3:2, 3:2, 4:3) fares when interactions between all disk particles are included.  Previous investigations \citep[e.g.,][]{Levison2011} have sought to approximate the excitation of the outer disk through different algorithms to mimic the viscous self-stirring.  \citeauthor{Levison2011} justified their approximation because fully active models were limited to artificially large bodies that result in enhanced numerical heating of the planetesimal disk \citep[e.g.,][]{Stewart2000}.  While \cite{Levison2011} included stirring due to close encounters between 1000 Pluto-mass bodies, their technique ignored long-range interactions between the bodies, which also significantly contributes to self-stirring \citep{Stewart2000}.  Moreover, these Pluto-mass bodies constituted $<$10\% of the total disk mass, and the remaining disk mass was not able to self-stir.  

\begin{figure}
\begin{center}
\includegraphics[width=\textwidth]{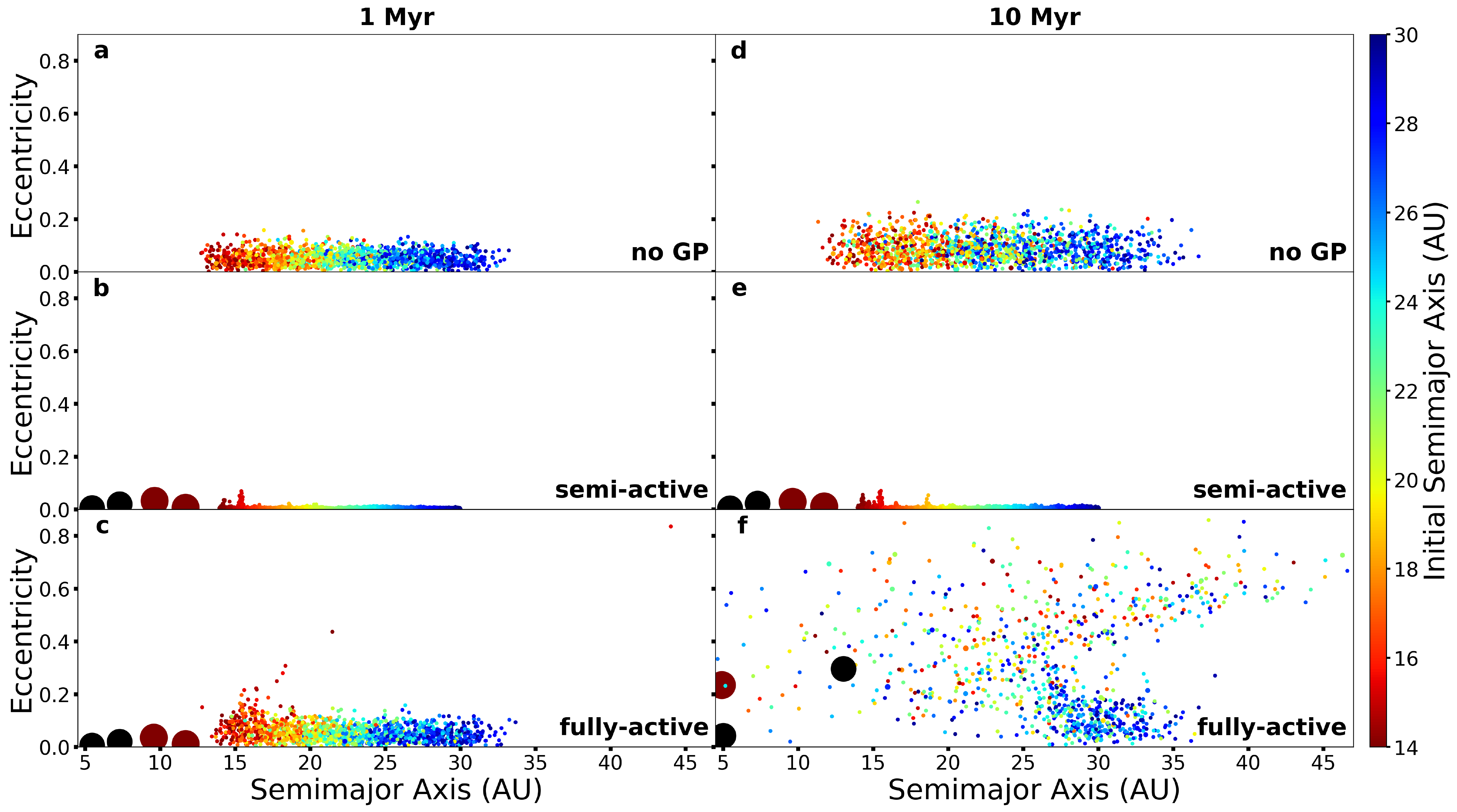}
\end{center}
\caption{Representations of the system state at 1 Myr (a -- c) and 10 Myr (d -- f) considering a Nice Model configuration with 4 giant planets with a 35 M$_\oplus$ outer disk ($a_i = 14$ AU) using either semi-active (b \& e) or fully-active particles (a, c, d, \& f).  The bottom panels (c \& f) use a similar initial disk of fully-active particles as the top panels (a \& d), where the giant planets are removed from the simulation.  The color-code represents the initial semimajor axis of the particles where the size of the points are scaled by the physical radius.}
\label{fig:modes}
\end{figure}

Within numerical software like \texttt{GENGA} and \texttt{mercury}, we have the option to ignore the interactions between disk particles, which we call the semi-active mode of the software and perform tests to examine the differences in the excitation of the disk. In Figure \ref{fig:modes}, we examine the eccentricity state as a function of semimajor axis at two epochs (1 \& 10 Myr) to better understand how the influence of the giant planets and the extent of mixing due to the disk excitation (using the color-scale).  Figs. \ref{fig:modes}a \& \ref{fig:modes}d show the full extent of the viscous self-stirring within an isolated disk (no giant planets) at these two different epochs.  Within 10 Myr, the maximum eccentricity grows to $\sim$0.2 allowing for significant portions of the outer disk to expand.  The stirring mixes the middle ($\sim$20-25 AU) portion of the disk and transports some bodies that originate near the inner edge to the outer edge.

Figs. \ref{fig:modes}b \& \ref{fig:modes}e show that mean motion resonances with the giant planets work to excite the inner portions ($<20$ AU) of the disk when using the semi-active mode, while the rest of the disk remains dynamically `cold'.  Also, the state of the 4 giant planets does not appreciably change on this timescale and the amount of disk spreading is minimal.  This is in contrast to Figs. \ref{fig:modes}a \& \ref{fig:modes}d that are much more excited despite the lack of giant planets in the simulations.  Also, our tests using the semi-active mode show that an isolated disk undergoes a negligible amount of excitation or spreading on these timescales.

Figs. \ref{fig:modes}c \& \ref{fig:modes}f show a considerably different scenario when the interactions between planetesimals are included.  In 1 Myr, disk bodies begin to cross orbits with the outer ice giant and induce its outward migration.  For this particular case, a resonance crossing between the giant planets ensues within 10 Myr, which is too quick to connect with the delayed instability in the classical Nice model \citep{Gomes2005}.  The interactions with the outer planetesimal disk are significant enough to trigger substantial migration of the outer ice giant and can lead to an overall instability of the giant planets on a 10 Myr timescale.  This may be a special case that depends on the disk gap $\Delta$ between the outer ice giant and the planetesimal disk, where a larger gap could delay the instability substantially \citep{Gomes2005}.

\begin{sidewaysfigure}
\begin{center}
\includegraphics[width=\textwidth]{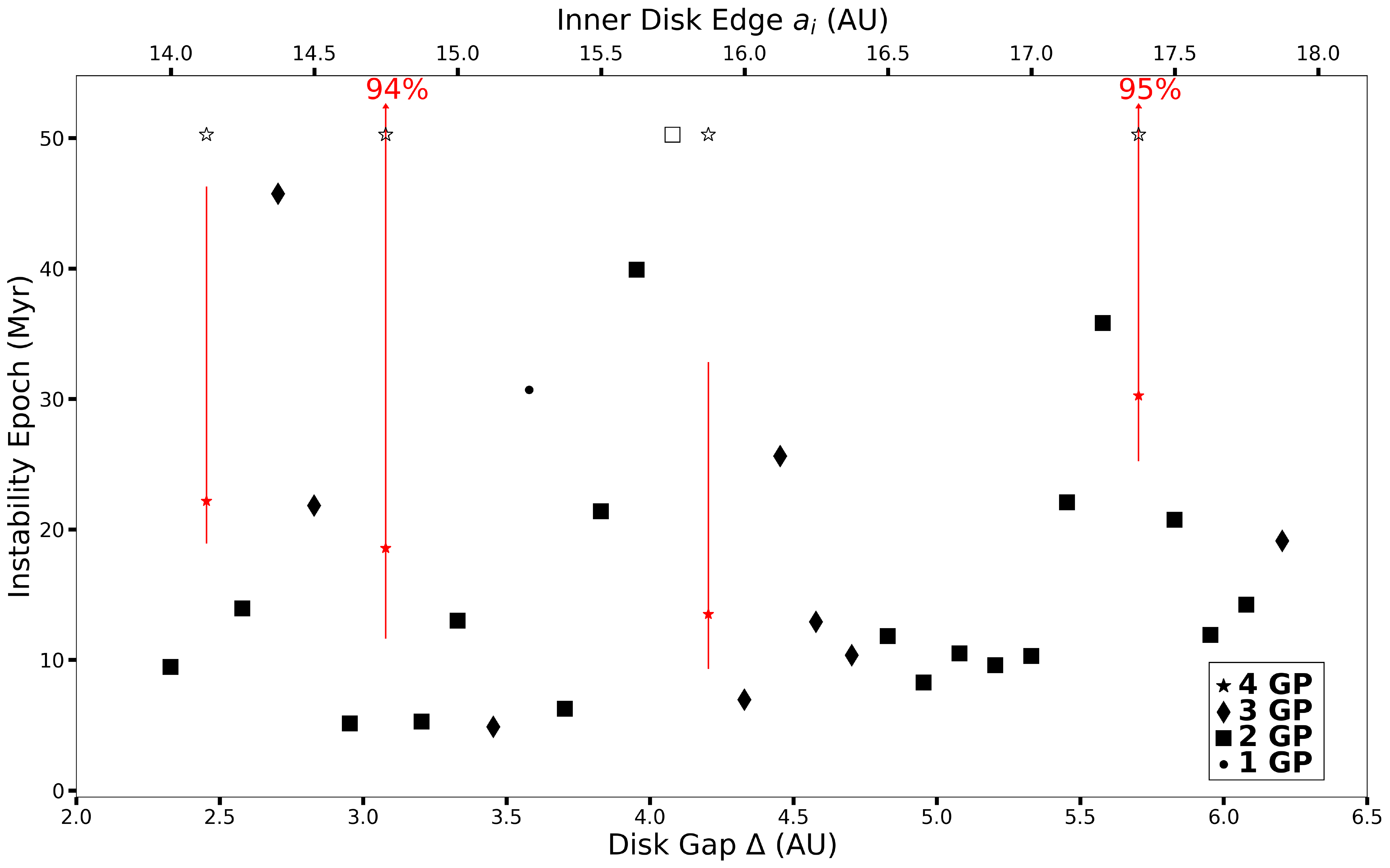}
\end{center}
\caption{Instability times of a Nice Model configuration with 4 giant planets along with a 35 M$_\oplus$ outer disk.  The solid black symbols indicate when a giant planet is lost, the open black stars denote that the configuration survives longer than 50 Myr, and the red symbols signify when 50\% of the disk is lost (accretion/ejection) and the errorbars represent the when either 16\% (lower bound) or 84\% of the disk is lost. Two of the runs ($a_i =$ 14.75 AU \& 17.375 AU) lose 84\% of the disk particles at $\sim$100 Myr and the percentages (in red) denote how much of the disk is lost at 550 Myr. }
\label{fig:life4}
\end{sidewaysfigure}

Thus, we perform a range of simulations that are given in Figure \ref{fig:life4} which shows the results as a function of the disk gap $\Delta$ systematically \citep[e.g.,][]{Levison2011} in order to uncover whether a delayed instability is possible or likely when considering our 4 giant planet resonant configuration.  The placement of symbols vertically in Fig. \ref{fig:life4} denotes when the first giant planet is ejected within a given simulation and the symbols themselves mark how many giant planets remain in the system up to the termination of the run.  There are two possibilities that can allow for Criterion $A$: a smooth migration of the outer ice giants that gradually depletes the outer disk or a scattering event that disrupts most of the outer disk but manages {to} retain all the giant planets. 

We find that both scenarios (smooth migration and scattering) are represented in our simulations and 4 out of 32 runs ($\sim$13\%) satisfy Criterion A.  However this typically includes a scattering event within 50 Myr and a smooth migration over the remaining 500 Myr.  In contrast to \cite{Levison2011}, our broader results show this to be a relatively uncommon occurrence, where $\sim$60\% of our simulations lost both ice giants and 20\% lost a single ice giant.  \cite{Nesvorny2012} also found that similar resonant configurations with low disk masses (M$_D < 50$ M$_\oplus$) typically lead to violent instabilities and planet ejection.  The mass of our disk particles may also play a role in allowing for early instabilities, but finding conditions that prevent planet ejection and replicate the outer solar system have been difficult for 4 giant planet configurations.  We have included points (in red) that mark when 50\% of the outer disk remains and errorbars that represent when either 16\% (lower bound) or 84\% (upper bound) of the disk is lost for cases where all 4 giant planets survive to 550 Myr.  Figure \ref{fig:life4} shows $\sim$16\% of our runs could be consistent with a late ($\sim$550 Myr) instability, thus we continue the runs that survived for 50 Myr for an order of magnitude longer in the simulation time to 550 Myr.  Only 1 of these simulations undergoes an instability on this longer timescale, where the other 4 simulations have smooth migrations of the outer giants.  The erosion of the disk is substantial and only $\sim3-4\%$ of the disk remains after the simulation ends at 550 Myr.

\begin{sidewaysfigure}
\begin{center}
\includegraphics[width=\textwidth]{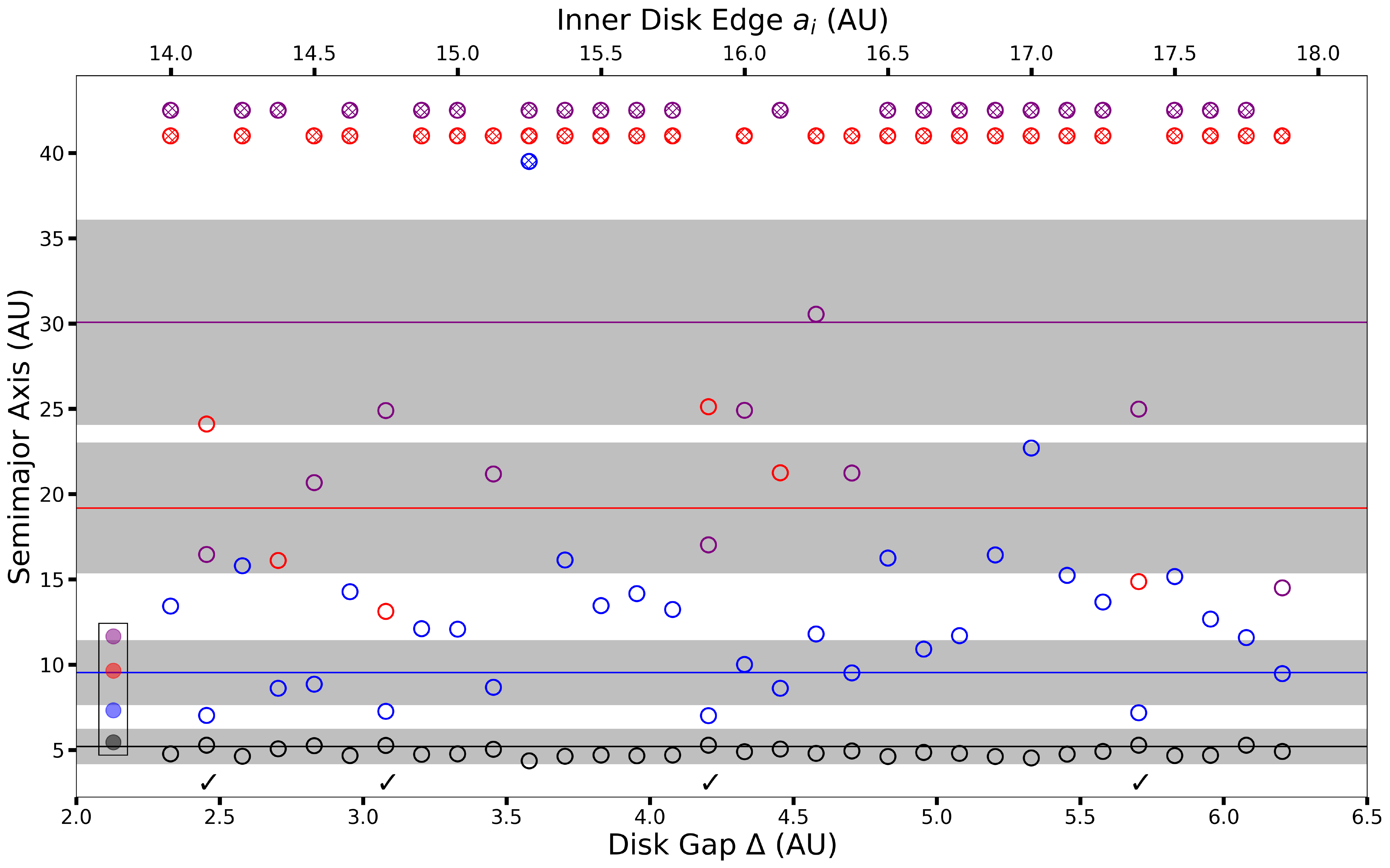}
\end{center}
\caption{Final giant planet architectures of a Nice Model configuration with 4 giant planets along with a 35 M$_\oplus$ outer disk.  The filled points enclosed in a box represent the initial giant planet configuration of each simulation, where the open points denote the final architectures.  Horizontal lines are given at the semimajor axis values of the current solar system and the gray bars represent the range of values within 20\%.  The checkmarks at the bottom identify those simulations that ended with 4 giant planets. }
\label{fig:Semi4}
\end{sidewaysfigure}

Figure \ref{fig:Semi4} illustrates the giant planet architectures in terms of the semimajor axes at the end of the simulations, where the initial configuration is shown color-coded in a box on the left, horizontal lines denote the current semimajor axes of the giant planets, and the 20\% range in gray that would accommodate Criterion B.  There are also checkmark (\cmark) symbols at the bottom, which mark cases where all four giant planets survived for 550 Myr.  Fig. \ref{fig:Semi4} shows that Criterion B is never satisfied as Jupiter and Saturn only leave the 3:2 mean motion resonance when a giant planet is lost and Criterion A becomes invalid.  The weaker condition, Criterion D, cannot be satisfied either, if Jupiter and Saturn remain in resonance.  After performing additional integrations for 10 Myr of these configurations, we find that Criterion C is satisfied in $\sim$70\% of the simulations showing the secular modes of the outer giants to be excited relatively easily by either the interactions with the outer disk or the loss of a giant planet.  Table \ref{tab:sum_4pl} summarizes our results and further delineates those cases that meet (\cmark) or fail (\xmark) each of our Criteria.

\subsection{5 Giant Planets} \label{sec:5GP}
Newer versions of the Nice Model have found that the ejection of an ice giant improves the chance of matching our success criteria \citep{Nesvorny2012,Deienno2017}.  One of the better configurations places the giant planets in a mean motion resonance with integer period ratios equal to 3:2, 3:2, 2:1, and 3:2.  We perform simulations similar to those presented in Section \ref{sec:4GP}, where the disk gap $\Delta$ varies in the same way (0.125 AU increments) in order to make fair comparisons between the two scenarios.  

\begin{figure}
\begin{center}
\includegraphics[width=\textwidth]{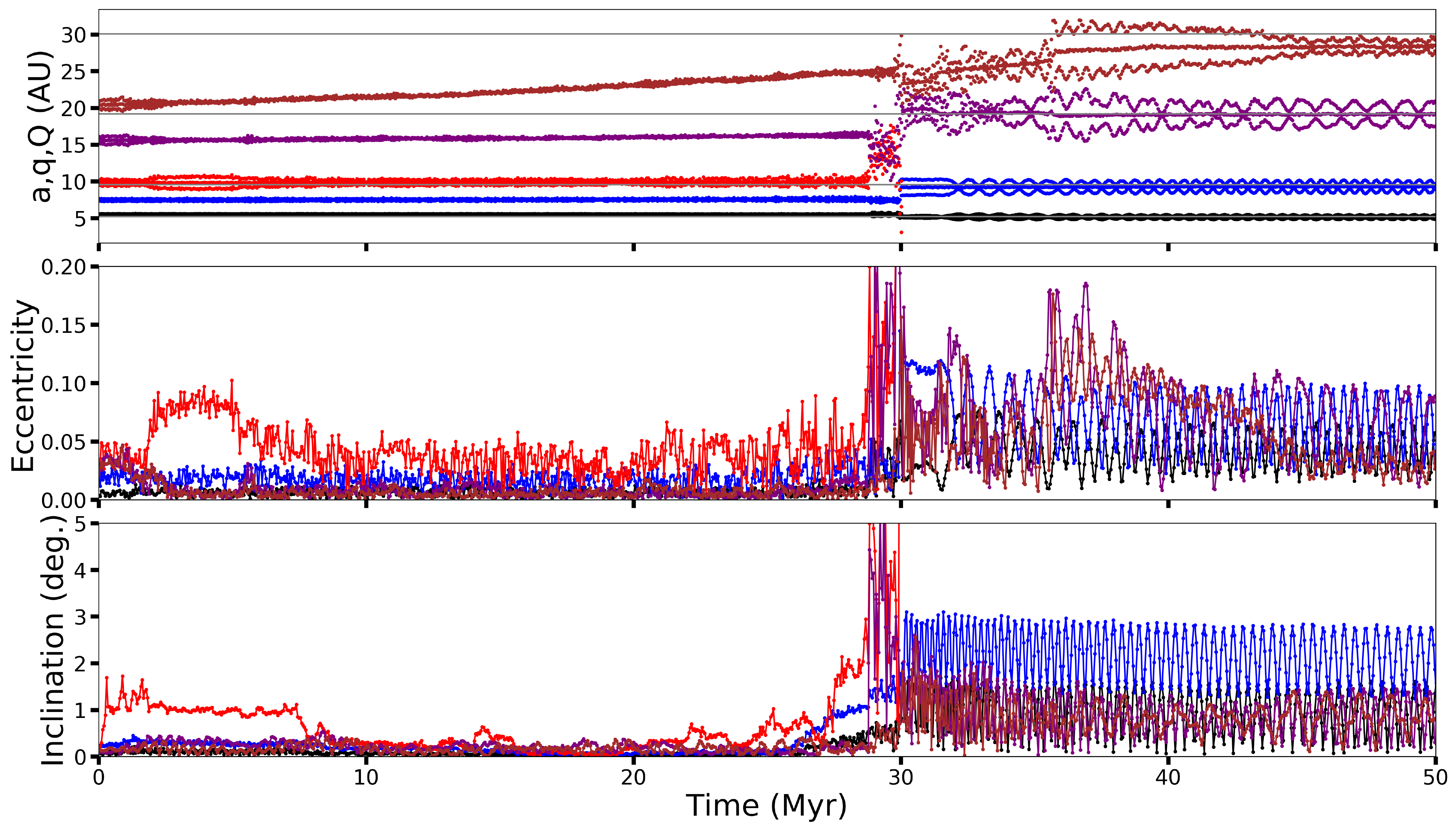}
\end{center}
\caption{Evolution of a successful configuration of 5 giant planets with a disk beginning at 26.197 AU ($\Delta = 5.829$ AU).  The evolution of the semimajor axis, periastron, and apastron (a,q,Q) are given (top panel) along with gray horizontal lines that represent the current semimajor axes of the modern giant planet configuration. The eccentricity (middle panel) and inclination (bottom panel) evolution are also shown to demonstrate the excitation due to scattering events and subsequent damping due to dynamical friction with the remains of the outer disk.}
\label{fig:Semi5_evol}
\end{figure}

Figure \ref{fig:Semi5_evol} shows the evolution of an example successful simulation ($\Delta = 5.829$ AU), where an instability of the giant planets occurs at $\sim$30 Myr.  This event substantially excites the innermost ice giant (red) allowing it to scatter off of Jupiter and escape from the system.  There is a secondary interaction a few million years later between the outer two ice giants, but this one is much more mild and allows the outermost ice giant to migrate outward and  closer to the present-day semimajor axis of Neptune.  There are 4 giant planets at the end of the simulation in Fig. \ref{fig:Semi5_evol}, which satisfies Criterion $A$.  Criterion B requires that the giant planets reside near their present-day semimajor axes, maintain average eccentricities $<$ 0.11, and average inclinations $<2^\circ$, which are both evident in Figure \ref{fig:Semi5_evol}.

\begin{figure}
\begin{center}
\includegraphics[width=\textwidth]{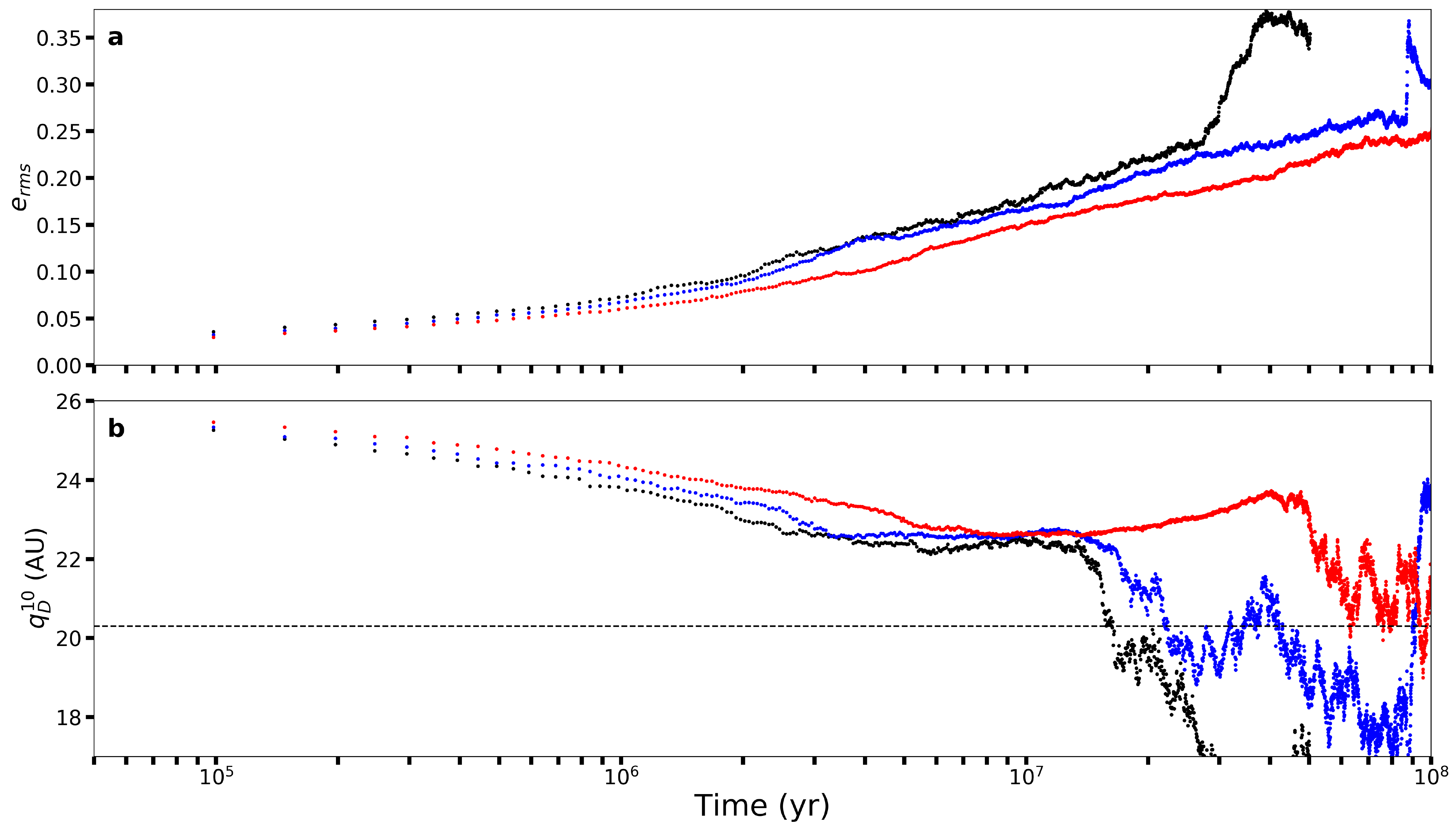}
\end{center}
\caption{{Evolution of the (a) rms eccentricity, and (b) the approximate borders of a disk ($a_i = 26.197$ AU) with 5 giant planets.  Each panel is color coded by the initial number of equal-mass particles: 1500 (black), 3000 (blue), and 4500 (red).  The run with 1500 particles is stopped at 50 Myr because the giant planet instability occurs after $\sim$30 Myr and the number of remaining particles is reduced significantly so that these statistical measures no longer apply.  Panel b delineates when the inner border of the disk interacts with the outermost giant planet (e.g., crossing the dashed line) using the 10$^{th}$ percentile in the periastron distribution as a proxy.}}
\label{fig:disk_5GP}
\end{figure}

{Figure \ref{fig:disk_5GP} shows the evolution of the disk in a similar way as in Figure \ref{fig:isodisk} (isolated disk), but with the 5 giant planets included and approximately double the surface density.  The different mass resolutions (1500, 3000, and 4500 particles) are color coded (black, blue, and red), respectively.  From Fig. \ref{fig:stir5}c we may expect higher eccentricities in Fig. \ref{fig:disk_5GP}a at 1 Myr, but the results in Fig. \ref{fig:stir5}c use a smaller disk gap resulting in a more significant initial perturbation from the outer giant planet and Fig. \ref{fig:stir5}a (red line) illustrates that the disk eccentricities are comparable to \ref{fig:disk_5GP}a at $2-4$ Myr.  Figure \ref{fig:disk_5GP}b shows diffusion of the disk through the 10$^{th}$ percentile in the periastron axis distribution.  The spreading of the disk causes orbit crossings with the outermost giant planet, where the timescale for these interactions depends on the mass resolution and only up to a factor of a few (i.e., less than an order of magnitude).}  

\begin{sidewaysfigure}
\begin{center}
\includegraphics[width=\textwidth]{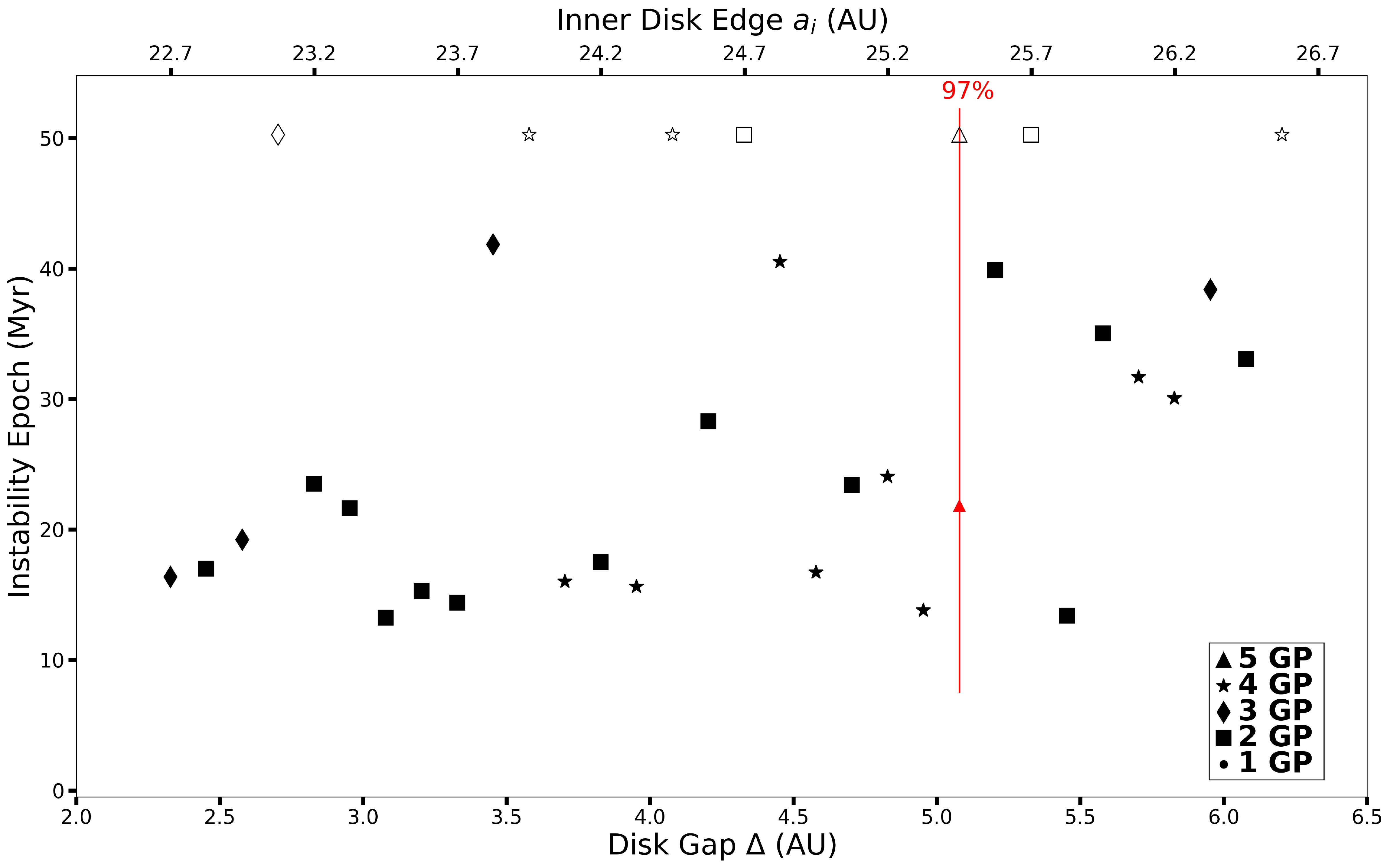}
\end{center}
\caption{Instability times of a Nice Model configuration with 5 giant planets along with a 20 M$_\oplus$ outer disk.  The black symbols indicate when a giant planet is lost, where the red symbols signify when 50\% of the disk is lost (accretion/ejection) and the errorbars represent the when either 16\% (lower bound) or 84\% of the disk is lost.  One of the runs ($a_i = 25.447 AU$) survives for 550 Myr and the percentage (in red) denotes how much of the disk is lost at the end of the simulation.}
\label{fig:life5}
\end{sidewaysfigure}

Our results in Figure \ref{fig:life5} indicate that 8 out of 32 runs (25\%) are able to scatter an ice giant out of the system within 50 Myr, while 3 runs (10\%) take a longer timescale to satisfy Criterion A.  To achieve this, it appears that the disk gap ($\Delta > 3.5$ AU) needs to be larger than was assumed in prior studies, where $\Delta \sim 1$ AU \citep{Gomes2005,Nesvorny2012}.  \cite{Levison2011} varied the disk gap for 4 planet systems and found that stable (1 Gyr) systems occurred once $\Delta>3.8$ AU.  \cite{Deienno2017} performed some tests varying $\Delta$ and found that $\Delta>2$ AU could be consistent with a late instability.  However they induced the migration of the outer ice giant with dust rather than encounters with planetesimals in the outer disk.  Our study finds that the timing of the instability typically occurs within 10--40 Myr without a strong trend in the disk gap $\Delta$.  Systems where all 5 giant planets are retained occur at a lower level ($\sim$3\%) than those we would deem to be a success, while most of the remaining simulations result in only 2 giant planets surviving the instability ($\sim$44\%).  We've included red points with errorbars to show that even though all 5 giant planets survive, a large portion (86\%) of the outer disk is lost within $\sim$50 Myr.  Over the longer timescale (550 Myr), the outer disk continues to erode until $97\%$ of the disk is lost.  The open symbols in Fig. \ref{fig:life5} denote runs that were unstable on timescales greater than 50 Myr, but less than 550 Myr.  These runs show that instabilities can occur fairly late, even with a small amount of disk material, where Table \ref{tab:sum_5pl} shows the instability times and whether each run meets our success Criteria. However, the minimal amount of disk material left also suggests that such late instabilities may be unable to generate the intense bombardment associated with the LHB, the original motivation for a late instability.

\begin{sidewaysfigure}
\begin{center}
\includegraphics[width=\textwidth]{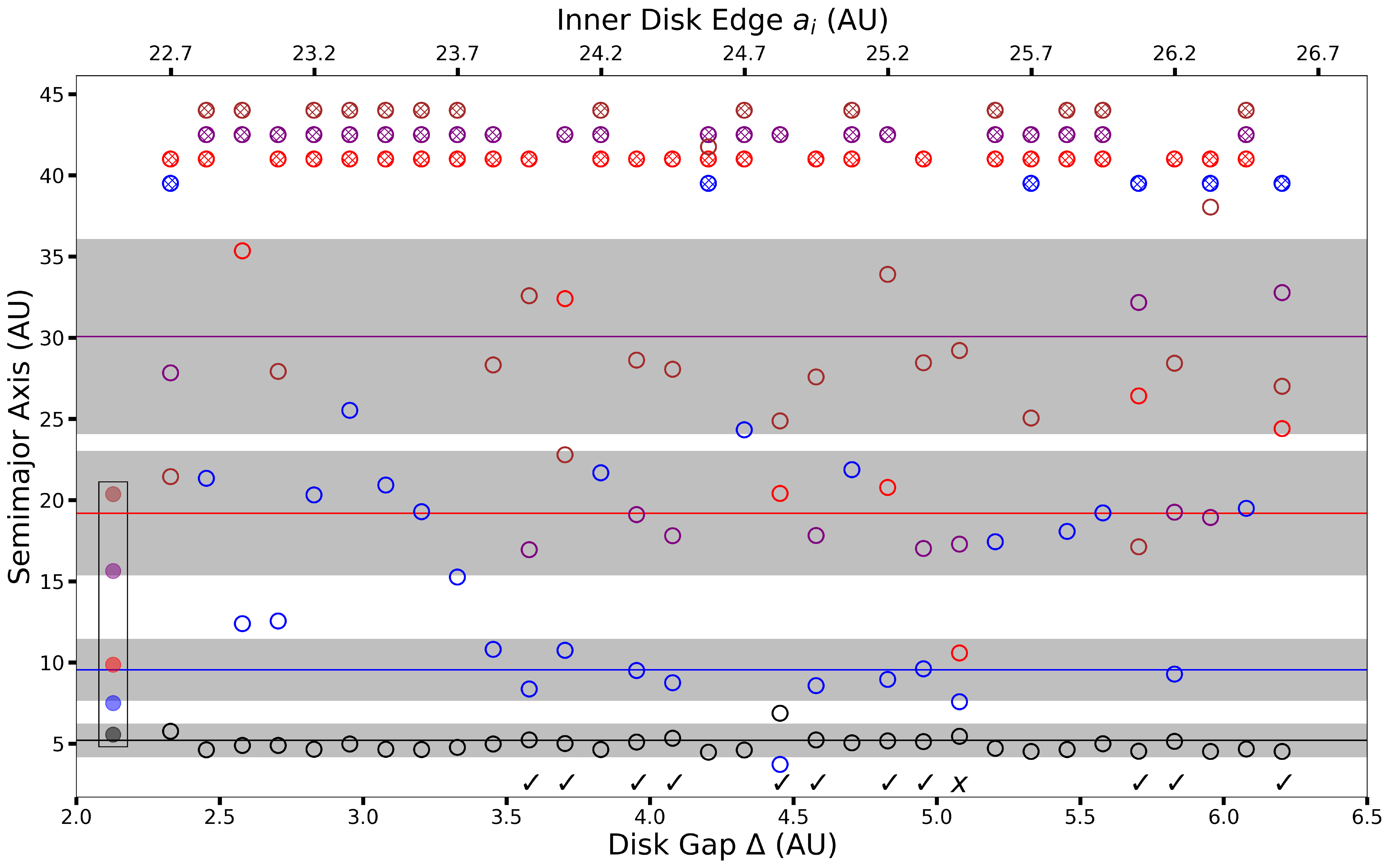}
\end{center}
\caption{Final giant planet architectures of a Nice Model configuration with 5 giant planets along with a 20 M$_\oplus$ outer disk.  The filled points enclosed in a box represent the initial giant planet configuration of each simulation, where the open points denote the final architectures.  Horizontal lines are given at the semimajor axis values of the current solar system and the gray bars represent the range of values within 20\%.  The checkmarks at the bottom identify those simulations that ended with 4 giant planets and the `X' marks show the simulations that did not lose any giant planets.  Points at the top filled with hatch marks denote which giant planets were lost in a given simulation. }
\label{fig:Semi5}
\end{sidewaysfigure}

We show the architectures in a similar way as in Section \ref{sec:4GP} in Figure \ref{fig:Semi5}, but mark the cases with 4 and 5 planets surviving with a checkmark (\cmark) and an ``$X$'', respectively.  From this view, we find that the run with 5 giant planets remaining did not allow for a large migration of Saturn or the inner ice giant, where the outer ice giant was able to substantially migrate outwards and begin depleting the outer disk.  The 8 cases that produced systems with 4 giant planets typically allow for them to arrive near the present-day semimajor axes.  There are 2 exceptions, where Saturn is either transported into the inner solar system or ejected entirely.  Fig. \ref{fig:Semi5} also shows that the next most common outcome (2 planets) typically leaves the system with only Jupiter and Saturn.

Table \ref{tab:sum_5pl} summarizes our results with respect to our success criteria.  Eleven of our simulations satisfy Criterion $A$, but only 4 of those also  satisfy Criterion B.  This is because the final mean eccentricity for these cases was larger than 0.11 indicating that the system becomes too dynamically `hot' to resemble the current solar system.  Similar to our results in Section \ref{sec:4GP}, Criterion $C$ is satisfied $\sim$68\% of the time.  One third of our runs that satisfy Criterion $A$, do not satisfy Criterion $D$.  However, satisfying Criterion $D$ is less of a concern because it was instituted to mitigate the strength of sweeping resonances from exciting the terrestrial region at a later epoch ($\sim$550 Myr).  Only 1 simulation satisfied all 4 Criteria, but this may change when we vary the total number of particles (see Section \ref{sec:num_part}).  Overall our results imply that a giant planet instability can occur within 50 Myr after the dispersal of the primordial gaseous disk, which overlaps with the late stages of accretion for the terrestrial planets, and largely resemble the current architecture of the giant planets.  Although we do not rule out the possibility of a delayed instability, these are not common in our simulation set and generally occur after the bulk of the disk mass has been dynamically depleted. 

\subsection{Dependance on the Number of Particles with Giant Planets} \label{sec:num_part}
\cite{Nesvorny2012}, \cite{ReyesRuiz2015}, and \cite{Deienno2017} evaluated whether the timing of the giant planet instability varied as a function of the number of bodies within the planetesimal disk.  We are motivated to do the same for this work and have performed simulations based upon a subset of configurations detailed in Sections \ref{sec:4GP} and \ref{sec:5GP}.  For the runs from Section \ref{sec:4GP}, we perform 8 runs beginning with $a_i=14$ AU and incrementing by 0.5 AU.  At each disk gap, the number of particles is changed by a factor of 0.5, 2, and 3, while keeping the initial disk mass constant (35 M$_\oplus$).  As a result, the mass of our disk particles range from several times the mass of the Moon down to a few times the mass of Pluto.

\begin{figure}
\begin{center}
\includegraphics[width=\textwidth]{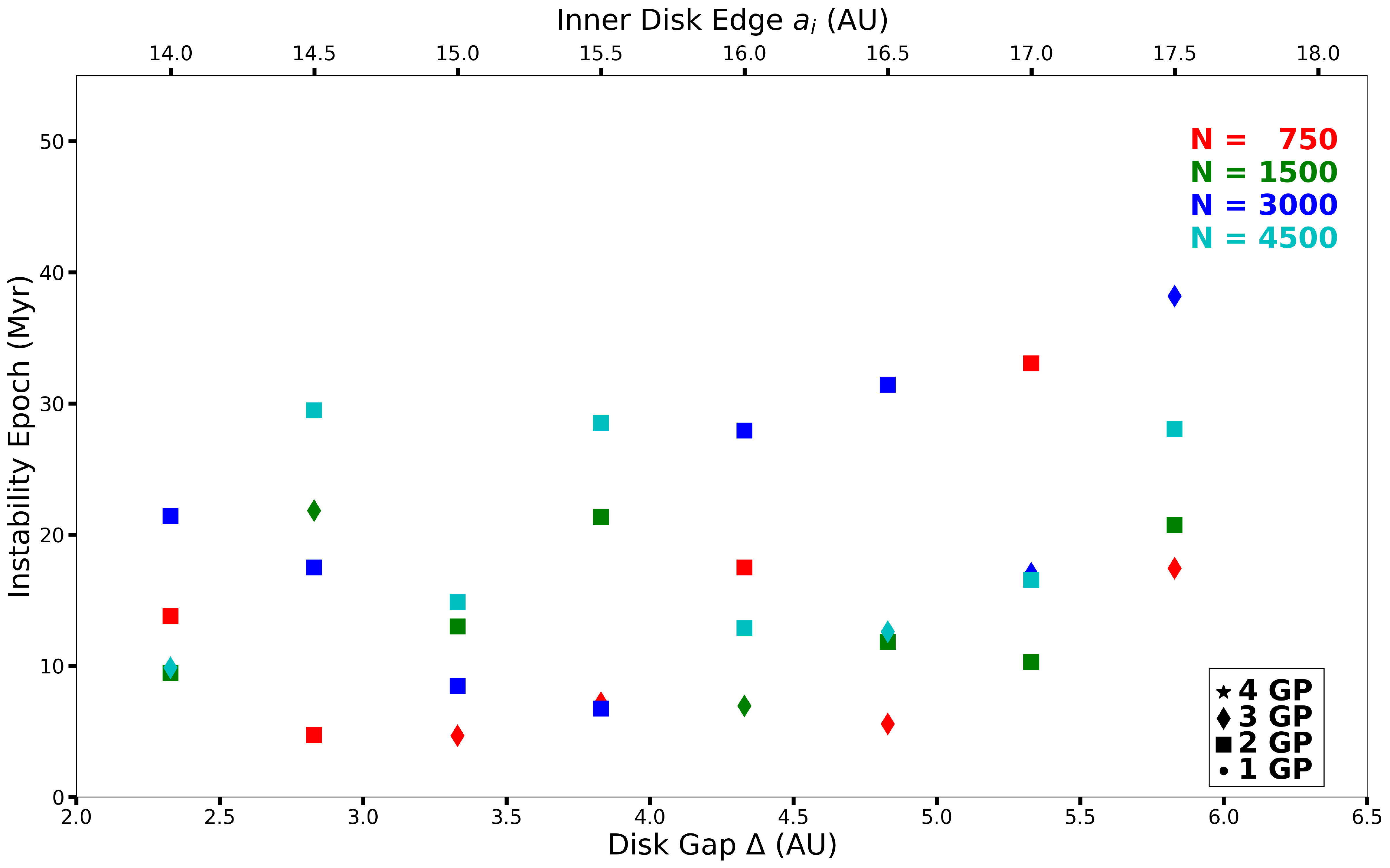}
\end{center}
\caption{Instability times of a Nice Model configuration with 4 giant planets along with a 35 M$_\oplus$ outer disk.  The symbols indicate how many giant planets remain at the end of the simulation and are color-coded by the initial number of disk particles.}
\label{fig:con_4pl}
\end{figure}

The results of these simulations are given in Figure \ref{fig:con_4pl}, where the points are color-coded relative to the initial number of particles within the outer planetesimal disk.  None of these cases produced a scenario consistent with a delayed instability.  The median instability epoch was 10.5, 14.1, 17.3, and 18.9  Myr after our simulations begin considering 750, 1500, 3000, and 4500 disk particles, respectively.  The scattering process is a chaotic one and our results show a large spread of actual outcomes relative to the number of giant planets remaining \cite[e.g.,][]{Kaib2016}.  Due to the small number of trials, we cannot rule out a delayed instability and can only infer that an early instability occurs more often irrespective of the number of disk particles.

\begin{figure}
\begin{center}
\includegraphics[width=\textwidth]{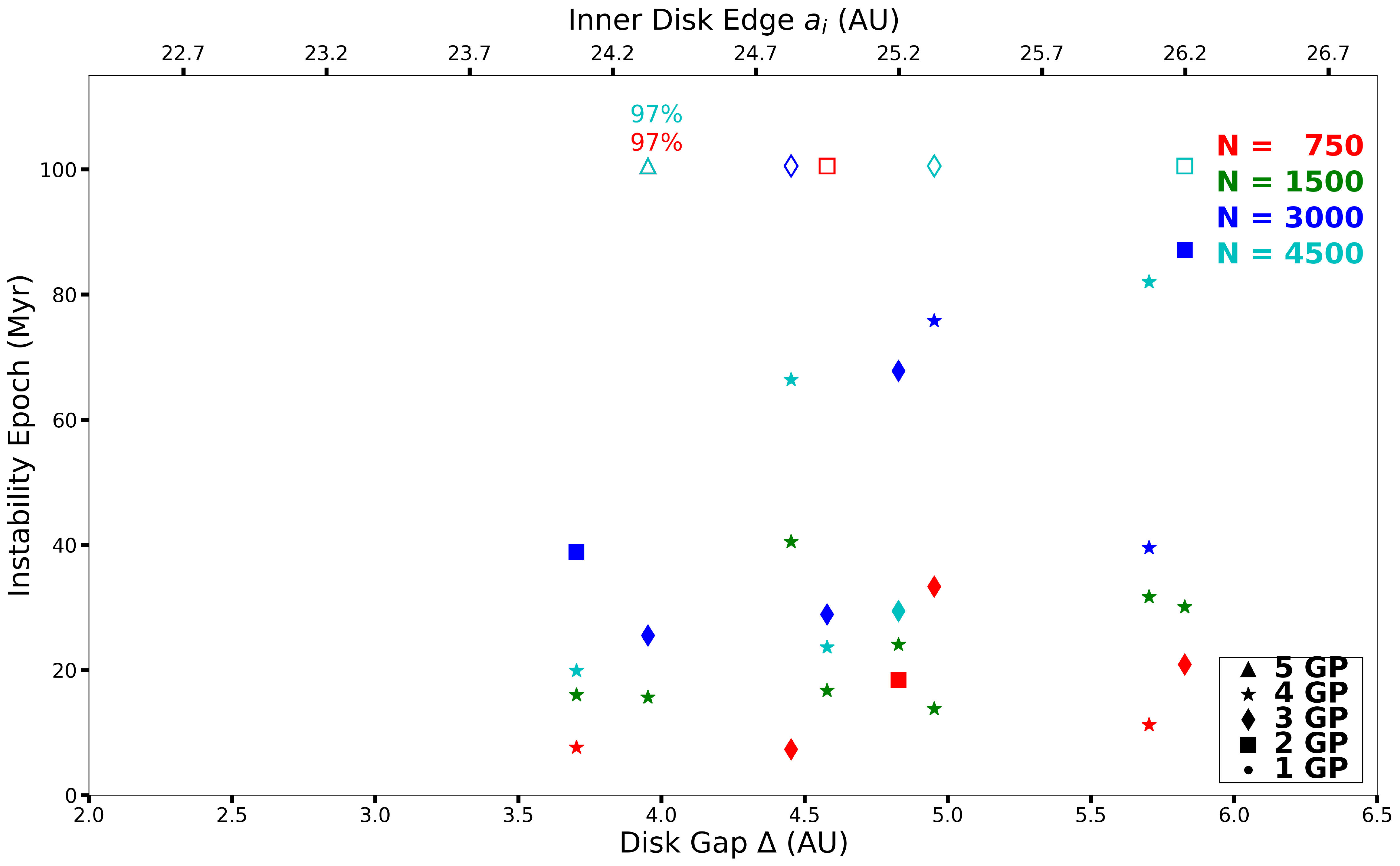}
\end{center}
\caption{Instability times of a Nice Model configuration with 5 giant planets along with a 20 M$_\oplus$ outer disk.  The symbols indicate how many giant planets remain at the end of the simulation and are color-coded by the initial number of disk particles.}
\label{fig:con_5pl}
\end{figure}

We perform another test using our eight 5-planet runs from Section \ref{sec:5GP} that underwent a giant planet instability within 50 Myr allowing 1 giant planet to escape and leaving 4 giant planets behind (i.e., filled stars in Fig. \ref{fig:life5}).  These simulations are not uniform relative to the disk gap.  These simulations vary the number of particles by a factor as before and keep the initial disk mass constant (20 M$_\oplus$), where the mass of the disk particles is 1.75x smaller than in our 4 giant planet runs.  The results of these simulations are given in Figure \ref{fig:con_5pl}, where the median instability epoch was 19.6, 26.2, 53.7, and 66.4 Myr after our simulations begin considering 750, 1500, 3000, and 4500 disk particles, respectively.

The median instability epoch is increasing for these simulations with increasing particle number, but the rate of increase is not dramatic.  {In terms of mean object mass, our 12,000 particle disk is comparable to the mean mass expected for the 12,000 most massive bodies in the primordial belt \citep{Shankman2013,Nesvorny2016}.  From Figure \ref{fig:isodisk}, we find that an isolated 4500 particle disk has an e-heating rate within 15\% of an isolated 12,000 particle disk and an a-spreading rate within 30 -- 40\% with an isolated disk without any planetary stirring.  When we include the giant planets, the mass resolution effects should be diminished further.  Finally, our higher resolution simulations only consider large disk gaps ($>$ 3.5 AU).  A larger initial disk gap would lower the immediate effects on the orbital evolution of the outermost giant planet due to the potential enhancement in $e_{rms}$ and a-spreading from the assumed mass resolution on 100 Myr timescales.  Based on this, we do not expect a still more realistic mass resolution will increase typical disk stability times by 1--2 orders of magnitude.  Moreover, numerical experiments probing such mass resolutions, beyond what we present here, exceed our current computing capabilities.} 

A majority of these additional runs resulted in giant planet instabilities on a timescale of 10 -- 40 Myr as before, where a subset (33\%) of these runs lasted for longer timescales.  In six of our runs (open symbols in Fig. \ref{fig:con_5pl}, two points overlap when $a_i = 24.322$ AU), 5 giant planets remained with a depleted disk after 100 Myr of simulation time.  We continue these runs to 550 Myr, a timescale consistent with a delayed instability.  Two simulations (750 and 4500 particles) where $\Delta \sim 4$ AU retained all 5 giant planets on this timescale and 97\% of the disk mass was ejected.  The other 4 runs underwent a giant planet instability on a timescale of $\sim$150 - 400 Myr and lost more than one giant planet.

\begin{figure}
\begin{center}
\includegraphics[width=\textwidth]{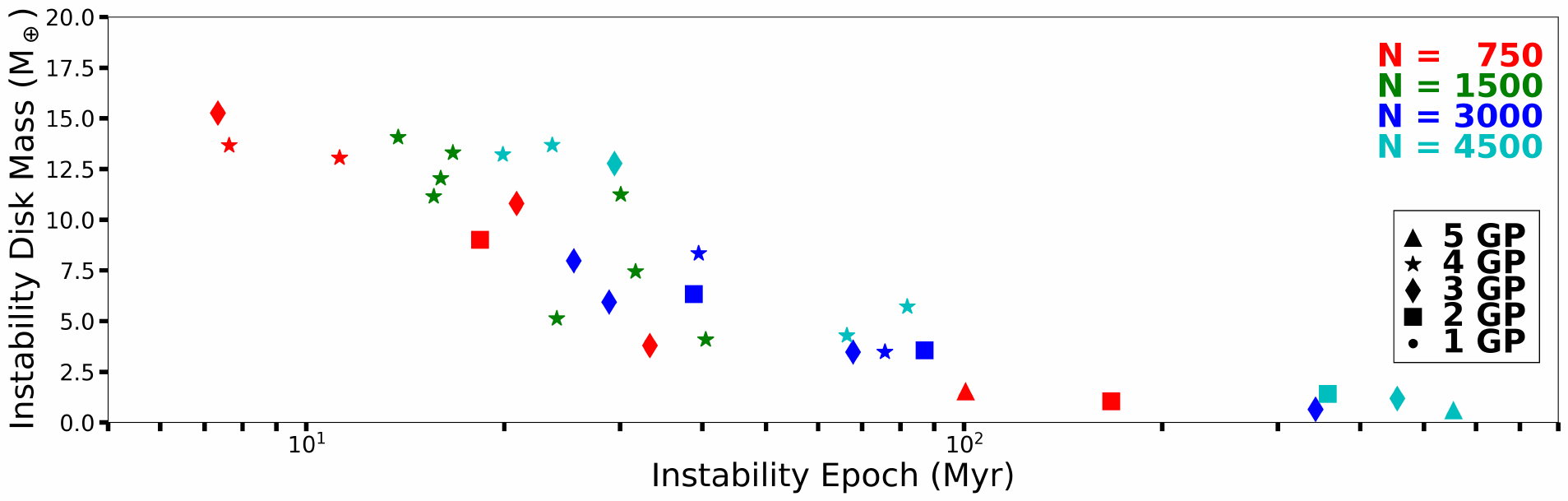}
\end{center}
\caption{Disk mass and instability times of a Nice Model configuration with 5 giant planets along with a 20 M$_\oplus$ outer disk.  The symbols indicate how many giant planets remain at the end of the simulation and are color-coded by the initial number of disk particles.}
\label{fig:instab_disk_5pl}
\end{figure}

Although only four runs have instabilities later than 150 Myr, a common feature among them is that their primordial Kuiper belts are all heavily depleted by the time of the instability.  Indeed, when we look at all of our 5 planet simulations we see that the instability time is strongly correlated with the amount of dynamical erosion of the disk prior to the onset of the instability. Moreover, this correlation appears to be largely independent of simulation resolution.  Figure \ref{fig:instab_disk_5pl} shows the disk mass (in M$_\oplus$) of our eight 5 planet simulations just prior to the instability time with respect to the initial number of particles.  We find that every simulation with an instability after 100 Myr has a disk mass of 2 M$_\oplus$ or less just prior to the instability.  This could have implications for the amount of material available for a LHB or whether a low disk mass could sufficiently damp the remaining giant planets' orbits after the instability, leading to another instability at a later time.  While the timing of the early instabilities tends to vary with the initial number of particles (within a factor of 2-3), later instabilities seem to always be associated with a heavily depleted disk, regardless of particle number.  

Having only four late instabilities makes it very difficult to assess the likelihood of the various orbital outcomes of late instabilities. We can further build our statistics by restarting these simulations right before they become unstable with slightly different conditions. The instability process is so chaotic that these slight shifts in conditions yield totally different final orbital configurations.  We probe how the outcome can vary with small perturbations on the innermost ice giant and how this affects our conclusions with respect to our success criteria.  

Small sets of simulations are performed using a state a few 100,000 yrs before the giant planets leave a compact configuration using the disk gaps and particle numbers shown in Fig. \ref{fig:con_5pl} (open symbols).   The perturbations are made by modifying the x-coordinate of the innermost ice giant randomly by 1 km and continuing the simulation for 10 Myr.  Table \ref{tab:per_5pl} shows the percentage of runs that satisfy each of our Criteria $A-D$ individually and simultaneously for short- and long-lived timescales.  We estimate our uncertainty in satisfying all 4 Criteria assuming binomial noise for $\sigma_{A-D}$.  These simulations suggest that the minority of systems that do experience a late instability meet our success criteria at a lower rate (6\%) than the systems which undergo early (t$<$100 Myr) instabilities (12\%).  

\subsection{Comparison with a Still Higher Mass Resolution} \label{sec:hires} 
{Until now, each of our simulations employs only one mass for every primordial belt object, and that mass is at least 2 Pluto Masses in most of our simulations. Meanwhile, KBO observations and formation models suggest that the primordial belt possessed 1000--4000 Pluto-mass objects and a much larger population of smaller objects \citep{Shankman2013,Nesvorny2016}.}  The number of particles necessary to accurately model the lower bounds of this constraint is $\sim$80,000 assuming a bimodal distribution consisting of 2000 Pluto-mass particles and the remainder in bodies that are 0.1 M$_P$ for a 20 M$_\oplus$ disk.  This is currently beyond our capability to simulate efficiently.

\begin{figure}
\begin{center}
\includegraphics[width=\textwidth]{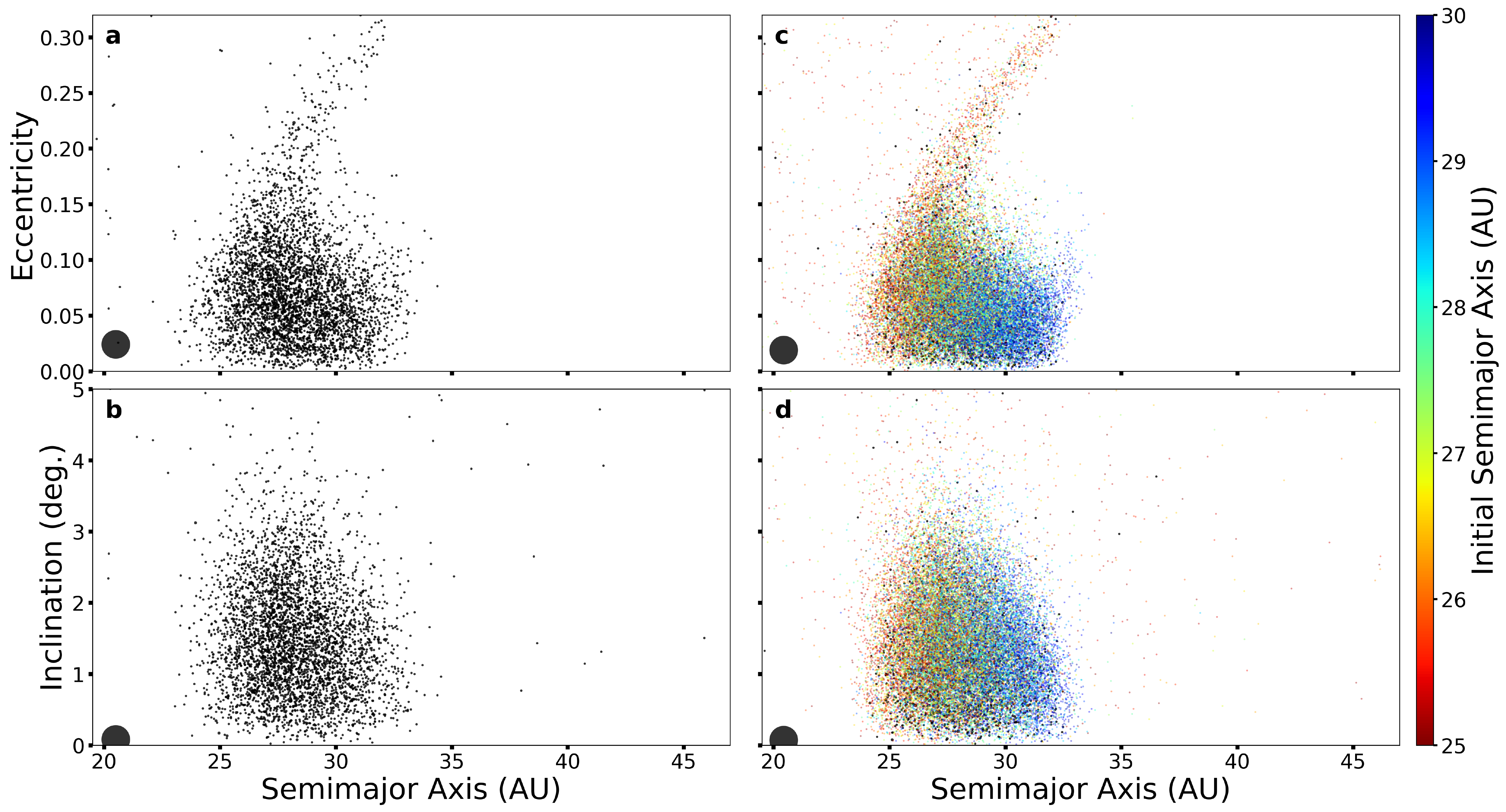}
\end{center}
\caption{Representations of the system state at 3 Myr considering a Nice Model configuration with 5 giant planets with a 20 M$_\oplus$ outer disk ($a_i \approx 25$ AU) using a disk of 4500 equal mass particles (a \& b) or using a bimodal distribution of 32,000 particles.  The color-code represents the initial semimajor axis of the semi-active particles where the size of the points are scaled by the physical radius.  The fully-active particles (black) are also scaled by the physical radius, but are not coded with the initial semimajor axis. }
\label{fig:con_5pl_hires}
\end{figure}

However, we can integrate a more realistic primordial Kuiper belt mass distribution for a modest time span and then compare this to our highest mass resolution, fully interacting runs to better gauge their dynamical accuracy. To do this, we utilize the semi-active mode of GENGA {for the large population of smaller bodies}.  We construct a 20 M$_\oplus$ disk that contains 40\% of its mass in 2000 fully active 2 M$_P$ bodies. The other 60\% of the disk mass is comprised of 30,000 bodies (each 0.2 M$_P$) that operate in the semi-active mode. While this disk's fully active bodies are still too massive by a factor of 2, the majority of the mass is in the semi-active form, which cannot self-stir, so these two effects will offset each other to some degree.  Interior to this highest resolution disk, we embed our 5 giant planet configuration, where the inner edge of the disk ($a_i$) begins at 25 AU so that we can compare with our run with $a_i = 24.947$ AU and 4500 particles.  This higher resolution simulation is stopped after 3 Myr so that our comparison primarily captures the relative magnitude of viscous stirring within the disk.  Figure \ref{fig:con_5pl_hires} illustrates the state of our higher resolution simulation (Figs. \ref{fig:con_5pl_hires}c \& \ref{fig:con_5pl_hires}d) with a comparable simulation with 4500 equal-mass particles (Figs. \ref{fig:con_5pl_hires}a \& \ref{fig:con_5pl_hires}b).  The black points represent those particles that are fully-active, where the color-coded points denote semi-active particles.  The eccentricity (Figs. \ref{fig:con_5pl_hires}a \& \ref{fig:con_5pl_hires}c) and inclination (Figs. \ref{fig:con_5pl_hires}b \& \ref{fig:con_5pl_hires}d) distributions appear strikingly similar on this timescale.

\begin{figure}
\begin{center}
\includegraphics[width=\textwidth]{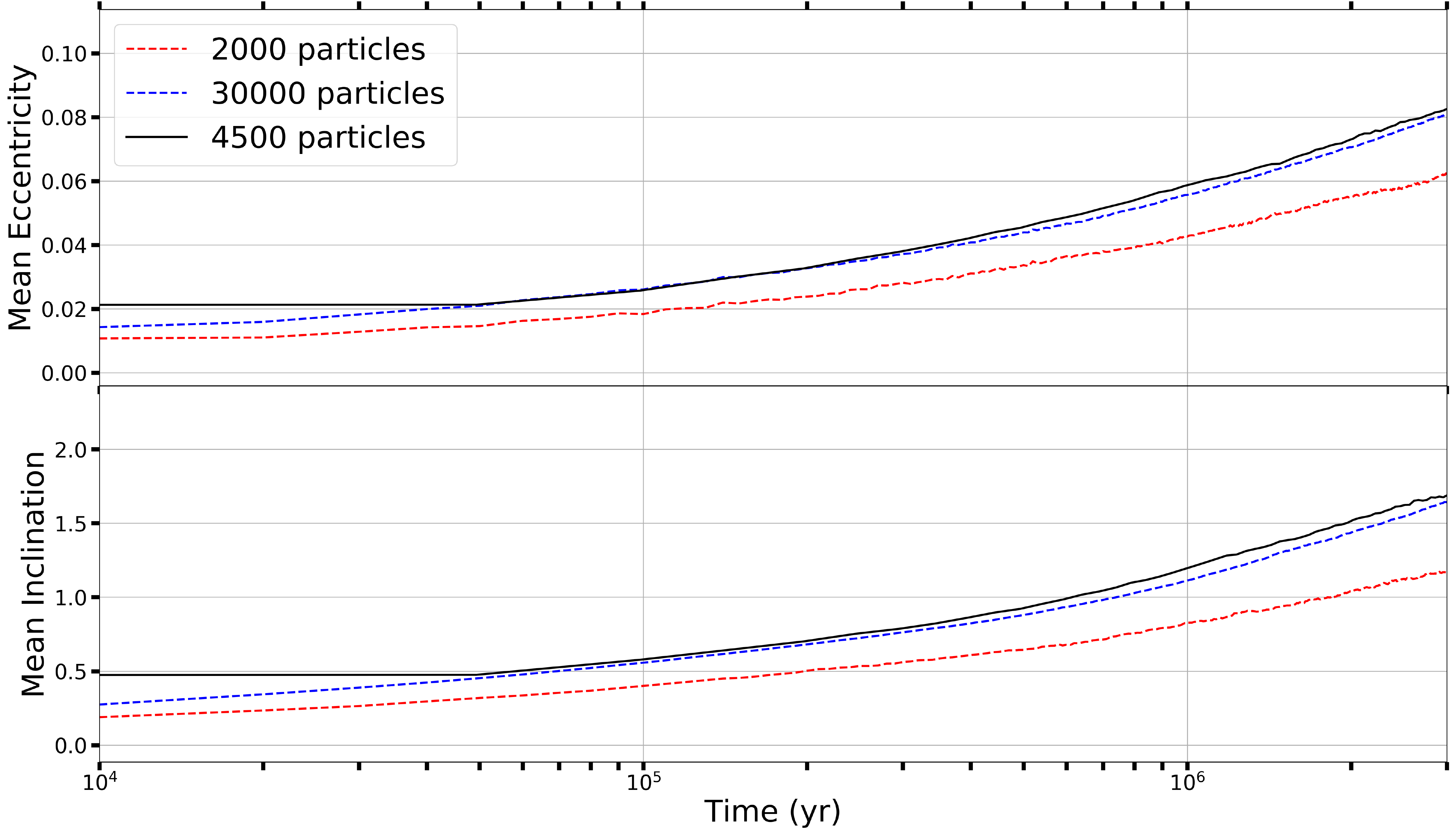}
\end{center}
\caption{Time evolution of the mass-weighted mean eccentricity (top) and inclination (bottom) of the outer disk particles for the simulations given in Figure \ref{fig:con_5pl_hires}.  The simulation with equal-masses (solid) is distinguished from the higher resolution run with a bimodal distribution (dashed), which is delineated further by the two mass bins (red \& blue) of 2 and 0.2 Pluto-mass bodies. }
\label{fig:con_5pl_dist}
\end{figure}

To distinguish between particle types, we provide the time evolution of the mass-weighted mean eccentricity and inclination of the disk particles in Figure \ref{fig:con_5pl_dist}.  We find that the smaller semi-active particles (dashed blue) closely track the evolution of the equal-mass particles (solid black) in both the mean eccentricity and inclination.  The fully-active particles within the high resolution run (dashed red) lag behind likely due to dynamical friction from the semi-active particles.  The total mass-weighted mean eccentricity of our most realistic disk will lie slightly closer to the blue line than the red, since 60\% of the disk mass is in the form of semi-active bodies. After 3 Myr of evolution, the mean eccentricities of the two disks are not radically different ($\sim$0.07 vs $\sim$0.08). Thus, we conclude that the real primordial Kuiper belt's dynamics and self-excitation should not be radically different from the highest mass-resolution of this work {on the timescales considered ($\sim$100s Myr)}.

\section{Conclusions} \label{sec:conc}
Overall, we find that disk self-stirring and spreading have a more significant effect on the orbital evolution of the giant planets than what has been previously assumed \citep{Levison2011,Nesvorny2012}.  {While many previous works have employed various approximations to handle the primordial disk's dynamics, we use GPUs to include the gravitational interactions between all disk bodies.  Our study uses both 4 and 5 giant planet configurations that begin in a mean motion resonance chain with an outer disk of planetesimals containing a total mass of 35 M$_\oplus$ and 20 M$_\oplus$, respectively.  In these simulations, the instability timescale of the giant planets in reaction to these disk forces occurs more often within 100 Myr from the start of the simulation.  However, to fully model belt particles' dynamics, our simulations must resort to using super-Pluto particle masses, which can artificially enhance the disk's self-stirring and accelerate the instability process.  An exploration of how the disk self-stirring depends on particle mass suggests that high-resolution (low particle-mass) belts will still allow long ($>$100 Myrs) delays before the onset of instability.  Thus, whether long instability times are possible depend on the properties of the solar system's protoplanetary disk.  If the disk's mass is primarily comprised of a single, heavy species, then we would expect it to evolve similarly to our low-resolution disks, tending toward short instability times. On the other hand, if the protoplanetary disk had a steep SFD or its mass was segregated into 2 or more species with large discrepancies in object mass, we'd expect instability times could be long, like those found in \cite{Levison2011}.}

In our simulations, the instability epoch with 4 giant planets appears to be independent of the assumed disk gap, or the distance between the inner edge of the disk and the outermost ice giant.  While some cases survived for 550 Myr, the amount of disk material was significantly reduced ($\sim 97\%$ M$_D$ lost) thereby leaving the trigger necessary for a delayed instability up to random perturbations between the giant planets and possibly reducing the disk's ability to damp the eccentricities of the giant planets if an instability does eventually occur.  Interestingly, we found the excitation of the Jupiter's eccentricity mode $e_{55}$ to proceed relatively easily compared to previous works that used approximations for the interactions within the disk \citep{Levison2011,Nesvorny2012}.  Previous Nice Model studies have generally disfavored Jupiter and Saturn beginning in the 2:1 resonance because this initial configuration fails to excite e55 enough, but the enhanced excitation seen in our work may reopen this possibility \citep{Nesvorny2012}. Future studies of this configuration using a self-stirring disk should address this.

Our 5 giant planet results showed a slight trend with the disk gap, favoring values $>$ 3.5 AU.  When we begin our 5-planet systems with a disks of 1500 equal mass particles, 1/8 of the post-instability systems reproduce the current orbits of the outer planets quite well (Criteria A, B, and D in Section \ref{sec:criteria}).  Seven out of 32 of these 5-planet systems remain stable for longer than 50 Myr. Upon further integration to 550 Myr, all but one of these systems went unstable.  The remaining stable system only retained 3\% of its disk mass after 550 Myr.  We note that the instabilities could occur at even later times ($\sim$800 Myr) like in the original models proposed by \cite{Gomes2005,Tsiganis2005}, but the mass flux needed to match a Late-Heavy Bombardment would likely not be available since the disk mass is so dynamically eroded by this point (i.e., all of the mass in our stable system would need to be directed at the inner solar system).

Given that disk particle masses of our main sets of simulations were several times the mass of Pluto, we also investigated how our systems' evolution varied with lower particle mass. For our 4-planet systems we do not find any significant trends with decreasing particle mass and increasing particle number.  However, in our 5 planet systems the median instability time increases to $\sim$66 Myr as the particle mass approaches 1 Pluto mass.  All of these additional runs assume a large gap between the disk and planets, and our highest resolution runs approach mean particle masses and numbers consistent with the primordial belt.  Even then, early instabilities within $\sim$20 Myr still occur, and the median instability time is well below 100 Myr.  We implemented statistical variations to our results using a random perturbation to the inner ice giant.  Following these cloned systems through instability indicated that the basic orbital architecture of the outer planets can be reproduced via late or early instabilities.  

Our systems with early instabilities successfully replicated the outer planets' orbits roughly 12\% of the time, while the late instability systems replicated the outer planets at the lower rate of 6\%.  This difference could be attributed in part to our resolution in disk particle masses.  However, comparing the self-stirring of a more realistic disk model that employs a bimodal set of KBO masses to that in our highest resolution self-interacting disks suggests that the disk dynamics of simulations will not radically change as we continue to approach more realistic mass resolutions.  Nevertheless, such higher mass resolution models will be required to confirm our conclusions with higher confidence.

\section*{Acknowledgements}

{The authors thank Kleomenis Tsiganis for a constructive and insightful review that greatly improved the clarity and quality of the manuscript.}   N.A.K. thanks the National Science Foundation for support under award AST-1615975 and the NASA Emerging Worlds Program for support under award 80NSSC18K0600.  The simulations presented here were performed using the OU Supercomputing Center for Education \& Research (OSCER) at the University of Oklahoma (OU).

\bibliographystyle{apj}
\bibliography{refs}

\newpage

\begin{deluxetable*}{ccccccc}
\tablecaption{ Summary of Results Considering a Nice Model Configuration with 4 Giant Planets\label{tab:sum_4pl}}

\tablehead{
\colhead{\hspace{.1cm}$a_i$}\hspace{.1cm} & \colhead{\hspace{.1cm}$\Delta$}\hspace{.1cm} & \colhead{\hspace{.1cm}$t$}\hspace{.1cm} & \colhead{\hspace{.1cm}$A$}\hspace{.1cm} & \colhead{\hspace{.1cm}$B$}\hspace{.1cm} & \colhead{\hspace{.1cm}$C$}\hspace{.1cm} & \colhead{\hspace{.1cm}$D$}\hspace{.1cm} \\
\colhead{\hspace{.1cm}(AU)}\hspace{.1cm} & \colhead{\hspace{.1cm}(AU)}\hspace{.1cm} & \colhead{\hspace{.1cm}(Myr)}\hspace{.1cm} &&&&} 
\startdata
14.000	&	2.329	&	9.462	&	\xmark	&	\xmark	&	\cmark	&	\xmark	\\
14.125	&	2.454	&	550.00	&	\cmark	&	\xmark	&	\cmark	&	\xmark	\\
14.250	&	2.579	&	13.947	&	\xmark	&	\xmark	&	\cmark	&	\xmark	\\
14.375	&	2.704	&	45.733	&	\xmark	&	\xmark	&	\xmark	&	\xmark	\\
14.500	&	2.829	&	21.832	&	\xmark	&	\xmark	&	\cmark	&	\xmark	\\
14.625	&	2.954	&	5.125	&	\xmark	&	\xmark	&	\cmark	&	\xmark	\\
14.750	&	3.079	&	550.00	&	\cmark	&	\xmark	&	\xmark	&	\xmark	\\
14.875	&	3.204	&	5.273	&	\xmark	&	\xmark	&	\cmark	&	\xmark	\\
15.000	&	3.329	&	13.010	&	\xmark	&	\xmark	&	\xmark	&	\xmark	\\
15.125	&	3.454	&	4.879	&	\xmark	&	\xmark	&	\xmark	&	\xmark	\\
15.250	&	3.579	&	30.702	&	\xmark	&	\xmark	&	\cmark	&	\xmark	\\
15.375	&	3.704	&	6.259	&	\xmark	&	\xmark	&	\cmark	&	\xmark	\\
15.500	&	3.829	&	21.388	&	\xmark	&	\xmark	&	\cmark	&	\xmark	\\
15.625	&	3.954	&	39.918	&	\xmark	&	\xmark	&	\cmark	&	\xmark	\\
15.750	&	4.079	&	119.90	&	\xmark	&	\xmark	&	\cmark	&	\xmark	\\
15.875	&	4.204	&	550.00	&	\cmark	&	\xmark	&	\xmark	&	\xmark	\\
16.000	&	4.329	&	6.949	&	\xmark	&	\xmark	&	\cmark	&	\cmark	\\
16.125	&	4.454	&	25.626	&	\xmark	&	\xmark	&	\cmark	&	\xmark	\\
16.250	&	4.579	&	12.912	&	\xmark	&	\xmark	&	\cmark	&	\xmark	\\
16.375	&	4.704	&	10.349	&	\xmark	&	\xmark	&	\xmark	&	\cmark	\\
16.500	&	4.829	&	11.828	&	\xmark	&	\xmark	&	\cmark	&	\xmark	\\
16.625	&	4.954	&	8.279	&	\xmark	&	\xmark	&	\cmark	&	\xmark	\\
16.750	&	5.079	&	10.497	&	\xmark	&	\xmark	&	\cmark	&	\xmark	\\
16.875	&	5.204	&	9.610	&	\xmark	&	\xmark	&	\cmark	&	\xmark	\\
17.000	&	5.329	&	10.300	&	\xmark	&	\xmark	&	\cmark	&	\xmark	\\
17.125	&	5.454	&	22.078	&	\xmark	&	\xmark	&	\cmark	&	\xmark	\\
17.250	&	5.579	&	35.828	&	\xmark	&	\xmark	&	\cmark	&	\xmark	\\
17.375	&	5.704	&	550.00	&	\cmark	&	\xmark	&	\cmark	&	\xmark	\\
17.500	&	5.829	&	20.747	&	\xmark	&	\xmark	&	\cmark	&	\xmark	\\
17.625	&	5.954	&	11.926	&	\xmark	&	\xmark	&	\xmark	&	\xmark	\\
17.750	&	6.079	&	14.242	&	\xmark	&	\xmark	&	\xmark	&	\xmark	\\
17.875	&	6.204	&	19.121	&	\xmark	&	\xmark	&	\xmark	&	\cmark	\\
\enddata
\tablecomments{Summary of results considering a Nice Model configuration (3:2,3:2,4:3) with 4 giant planets ($a_4^{GP}$ $\approx$ 11.6 AU) along with a 35 M$_\oplus$ outer disk. The columns correspond to the heliocentric inner edge of the disk $a_i$, the distance between the inner disk edge with the outer ice giant $\Delta$, the time of the giant planet instability $t$, and whether the given conditions meet (\cmark) or fail (\xmark) each of the our criteria for success $A-D$.  }

\end{deluxetable*}

\newpage

\begin{deluxetable*}{ccccccc}
\tablecaption{Summary of Results Considering a Nice Model Configuration with 5 Giant Planets \label{tab:sum_5pl}}
\tablehead{
\colhead{\hspace{.1cm}$a_i$}\hspace{.1cm} & \colhead{\hspace{.1cm}$\Delta$}\hspace{.1cm} & \colhead{\hspace{.1cm}$t$}\hspace{.1cm} & \colhead{\hspace{.1cm}$A$}\hspace{.1cm} & \colhead{\hspace{.1cm}$B$}\hspace{.1cm} & \colhead{\hspace{.1cm}$C$}\hspace{.1cm} & \colhead{\hspace{.1cm}$D$}\hspace{.1cm} \\
\colhead{\hspace{.1cm}(AU)}\hspace{.1cm} & \colhead{\hspace{.1cm}(AU)}\hspace{.1cm} & \colhead{\hspace{.1cm}(Myr)}\hspace{.1cm} &&&&} 
\startdata
%
22.697	&	2.329	&	16.361	&	\xmark	&	\xmark	&	\xmark	&	\xmark	\\
22.822	&	2.454	&	17.002	&	\xmark	&	\xmark	&	\cmark	&	\xmark	\\
22.947	&	2.579	&	19.220	&	\xmark	&	\xmark	&	\cmark	&	\xmark	\\
23.072	&	2.704	&	56.969	&	\xmark	&	\xmark	&	\cmark	&	\xmark	\\
23.197	&	2.829	&	23.507	&	\xmark	&	\xmark	&	\cmark	&	\xmark	\\
23.322	&	2.954	&	21.635	&	\xmark	&	\xmark	&	\cmark	&	\xmark	\\
23.447	&	3.079	&	13.257	&	\xmark	&	\xmark	&	\cmark	&	\xmark	\\
23.572	&	3.204	&	15.277	&	\xmark	&	\xmark	&	\cmark	&	\xmark	\\
23.697	&	3.329	&	14.390	&	\xmark	&	\xmark	&	\cmark	&	\xmark	\\
23.822	&	3.454	&	41.840	&	\xmark	&	\xmark	&	\cmark	&	\xmark	\\
23.947	&	3.579	&	288.83	&	\cmark	&	\xmark	&	\xmark	&	\cmark	\\
24.072	&	3.704	&	16.016	&	\cmark	&	\xmark	&	\cmark	&	\cmark	\\
24.197	&	3.829	&	17.495	&	\xmark	&	\xmark	&	\cmark	&	\xmark	\\
24.322	&	3.954	&	15.622	&	\cmark	&	\xmark	&	\cmark	&	\cmark	\\
24.447	&	4.079	&	378.33	&	\cmark	&	\cmark	&	\xmark	&	\cmark	\\
24.572	&	4.204	&	28.287	&	\xmark	&	\xmark	&	\cmark	&	\xmark	\\
24.697	&	4.329	&	133.16	&	\xmark	&	\xmark	&	\cmark	&	\xmark	\\
24.822	&	4.454	&	40.509	&	\cmark	&	\xmark	&	\xmark	&	\xmark	\\
24.947	&	4.579	&	16.706	&	\cmark	&	\cmark	&	\cmark	&	\cmark	\\
25.072	&	4.704	&	23.409	&	\xmark	&	\xmark	&	\cmark	&	\cmark	\\
25.197	&	4.829	&	24.049	&	\cmark	&	\xmark	&	\xmark	&	\cmark	\\
25.322	&	4.954	&	13.799	&	\cmark	&	\cmark	&	\xmark	&	\cmark	\\
25.447	&	5.079	&	550.00	&	\xmark	&	\xmark	&	\xmark	&	\xmark	\\
25.572	&	5.204	&	39.869	&	\xmark	&	\xmark	&	\cmark	&	\xmark	\\
25.697	&	5.329	&	70.423	&	\xmark	&	\xmark	&	\cmark	&	\xmark	\\
25.822	&	5.454	&	13.405	&	\xmark	&	\xmark	&	\cmark	&	\xmark	\\
25.947	&	5.579	&	35.039	&	\xmark	&	\xmark	&	\cmark	&	\xmark	\\
26.072	&	5.704	&	31.688	&	\cmark	&	\xmark	&	\xmark	&	\xmark	\\
26.197	&	5.829	&	30.062	&	\cmark	&	\cmark	&	\xmark	&	\cmark	\\
26.322	&	5.954	&	38.390	&	\xmark	&	\xmark	&	\cmark	&	\xmark	\\
26.447	&	6.079	&	33.067	&	\xmark	&	\xmark	&	\cmark	&	\xmark	\\
26.572	&	6.204	&	89.988	&	\cmark	&	\xmark	&	\xmark	&	\xmark	\\
\enddata
\tablecomments{Summary of results considering a Nice Model configuration (3:2,3:2,2:1,3:2) with 5 giant planets ($a_5^{GP}$ $\approx$ 20.3 AU) along with a 20 M$_\oplus$ outer disk. The columns correspond to the heliocentric inner edge of the disk $a_i$, the distance between the inner disk edge with the outer ice giant $\Delta$, the time of the giant planet instability $t$ in Myr, and whether the given conditions meet (\cmark) or fail (\xmark) each of the our criteria for success $A-D$. } 

\end{deluxetable*}

\begin{deluxetable}{ccccccccc}
\tablecaption{Statistical Outcomes on Instability Time for a 5 Giant Planet Configuration \label{tab:per_5pl}}
\tablecolumns{9}
\tablehead{
 \colhead{$t$} &  \colhead{$A$} & \colhead{$B$} & \colhead{$C$} & \colhead{$D$} & \colhead{$A-D$} & \colhead{$\sigma_{A-D}$} \\
 \colhead{(Myr)} & \colhead{(\%)} &\colhead{(\%)}&\colhead{(\%)}&\colhead{(\%)}&\colhead{(\%)}&\colhead{(\%)}} 
\startdata
$<$100	&	30\%	&	20\%	&	98\%	&	30\%	&	12\%	&	3\%	 \\
$>$100	&	28\%	&	11\%	&	64\%	&	29\%	&	6\%	&	2\%	 \\
\enddata
\tablecomments{Statistical outcomes (120 simulations each) due to a small perturbation considering the short- ($<100$ Myr) and long-lived ($>$100 Myr) 5 giant planet configurations. The columns correspond to the temporal range of the giant planet instability $t$ in Myr, the percentage of runs that satisfy Criteria $A-D$ \textit{individually}, the percentage of runs that satisfy Criteria $A-D$ \textit{simultaneously}, and the uncertainty in our estimation of satisfying all 4 Criteria simultaneously. }
\end{deluxetable}

\end{document}